\documentclass[conference]{IEEEtran}
\IEEEoverridecommandlockouts
\usepackage{cite}
\usepackage{amsmath,amssymb,amsfonts}
\usepackage{graphicx}
\usepackage{textcomp}
\usepackage{xcolor}
\def\BibTeX{{\rm B\kern-.05em{\sc i\kern-.025em b}\kern-.08em
    T\kern-.1667em\lower.7ex\hbox{E}\kern-.125emX}}

\usepackage{enumitem}

\usepackage{graphicx}
\usepackage{textcomp}
\usepackage{xcolor}

\usepackage{url,caption,subcaption,graphicx}
\usepackage{tabularx,soul}

\usepackage{array}
\newcolumntype{P}[1]{>{\centering\arraybackslash}p{#1}}
\newcolumntype{M}[1]{>{\centering\arraybackslash}m{#1}}

\usepackage{textcomp}
\usepackage{pgfplots}
\usepackage{tcolorbox}

\pgfplotsset{width=8.4cm,height=6cm,compat=1.17}
\usepackage{amsmath,xcolor,soul, fontawesome,algpseudocode,xspace,multirow,ctable,longtable,colortbl,lscape,pifont}

\sethlcolor{red}
\usepackage[utf8]{inputenc}
\usepackage[font={small}]{caption}
\usepackage{listings}
\usepackage{xcolor}
\usepackage{pifont}
\usepackage{tikz}

\definecolor{codegreen}{rgb}{0,0.6,0}
\definecolor{codegray}{rgb}{0.5,0.5,0.5}
\definecolor{codepurple}{rgb}{0.58,0,0.82}
\definecolor{backcolour}{rgb}{0.95,0.95,0.92}
\definecolor{bblue}{HTML}{4F81BD}
\definecolor{rred}{HTML}{E11916}
\definecolor{ggreen}{HTML}{3FD72D}
\definecolor{ggreen1}{HTML}{9DEC9D}
\definecolor{ppurple}{HTML}{9F4C7C}
\definecolor{yyellow}{HTML}{FFC000}
\definecolor{yyellow1}{HTML}{FEE599}

\definecolor{debug}{HTML}{FFBABA}
\definecolor{info}{HTML}{FF5252}
\definecolor{warning}{HTML}{FF0000}
\definecolor{severe}{HTML}{A70000}

\definecolor{last-year}{HTML}{1E476C}
\definecolor{this-year}{HTML}{9FC5E8}

\lstdefinestyle{mystyle}{
  backgroundcolor=\color{backcolour},   commentstyle=\color{codegreen},
  keywordstyle=\color{magenta},
  numberstyle=\tiny\color{codegray},
  stringstyle=\color{codepurple},
  basicstyle=\ttfamily\footnotesize,
  breakatwhitespace=false,         
  breaklines=true,                 
  captionpos=b,                    
  keepspaces=true,                 
  numbers=left,                    
  numbersep=5pt,                  
  showspaces=false,                
  showstringspaces=false,
  showtabs=false,                  
  tabsize=1 
}

\usepackage{booktabs,siunitx,comment}

\newcommand{\ie}{\emph{i.e.,}\xspace}
\newcommand{\eg}{\emph{e.g.,}\xspace}

\newcommand{\rwa}{{$R_{with-ads}$}\xspace}
\newcommand{\rwoa}{{$R_{without-ads}$}\xspace}
\newcommand{\roa}{{$R_{only-ads}$}\xspace}

\usepackage{picture} 
\usepackage{cite}
\usepackage{booktabs} 

\begin{document}
\title{Accessibility Issues in Ad-Driven Web Applications}

\author{\IEEEauthorblockN{Abdul Haddi Amjad}
\IEEEauthorblockA{\textit{Computer Science} \\
\textit{Virginia Tech}\\
Blacksburg, USA \\
hadiamjad@vt.edu}
\and
\IEEEauthorblockN{Muhammad Danish}
\IEEEauthorblockA{\textit{Computer Science} \\
\textit{Virginia Tech}\\
Blacksburg, USA \\
mdanish@vt.edu}
\and
\IEEEauthorblockN{Bless Jah}
\IEEEauthorblockA{\textit{Computer Science} \\
\textit{Virginia Tech}\\
Blacksburg, USA \\
blessj@vt.edu}
\and
\IEEEauthorblockN{Muhammad Ali Gulzar}
\IEEEauthorblockA{\textit{Computer Science} \\
\textit{Virginia Tech}\\
Blacksburg, USA \\
gulzar@cs.vt.edu}
}

\maketitle
\begin{abstract}
Website accessibility is essential for inclusiveness and regulatory compliance. Although third-party advertisements (ads) are a vital revenue source for free web services, they introduce significant accessibility challenges. Leasing a website's space to ad-serving technologies like DoubleClick results in developers losing control over ad content accessibility. Even on highly accessible websites, third-party ads can undermine adherence to Web Content Accessibility Guidelines (WCAG).
We conduct the first-of-its-kind large-scale investigation of 430K website elements, including nearly 100K ad elements, to understand the accessibility of ads on websites. 
We seek to understand the prevalence of inaccessible ads and their overall impact on the accessibility of websites.
Our findings show that 67\% of websites experience increased accessibility violations due to ads, with common violations including Focus Visible (WCAG 2.4.7), and On Input (WCAG 3.2.2). 
Popular ad-serving technologies like Taboola, DoubleClick, and RevContent often serve ads that fail to comply with WCAG standards. 
Even when ads are WCAG compliant, 27\% of them have alternative text in ad images that misrepresents information, potentially deceiving users. 
Manual inspection of a sample of these misleading ads revealed that user-identifiable data is collected on 94\% of websites through interactions, such as hovering or pressing enter.
%
Since users with disabilities often rely on tools like screen readers that require hover events to access website content, they have no choice but to compromise their privacy in order to navigate website ads. 
Based on our findings, we further dissect the root cause of these violations and provide design guidelines to both website developers and ad-serving technologies to achieve WCAG-compliant ad integration.

\end{abstract}

\begin{IEEEkeywords}
accessibility, web, ads, privacy
\end{IEEEkeywords}

\section{Introduction}
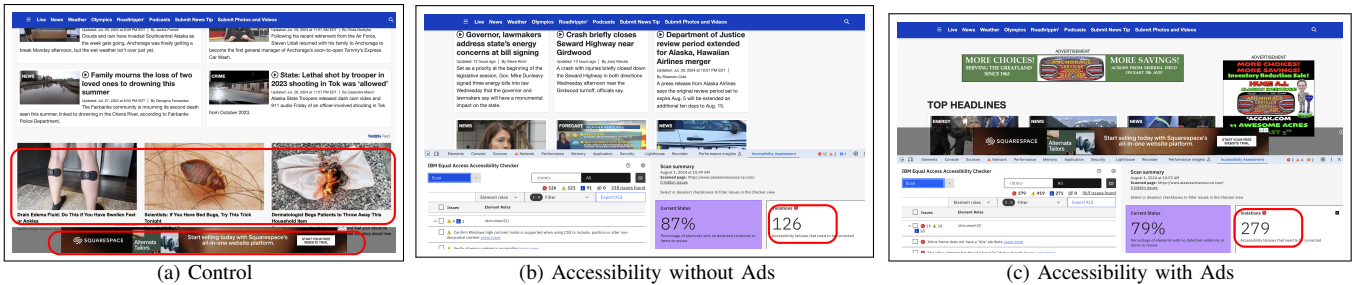
\begin{figure*}[!t]
\centering
\begin{subfigure}{.295\textwidth}
    \fbox{\includegraphics[width=.95\linewidth]{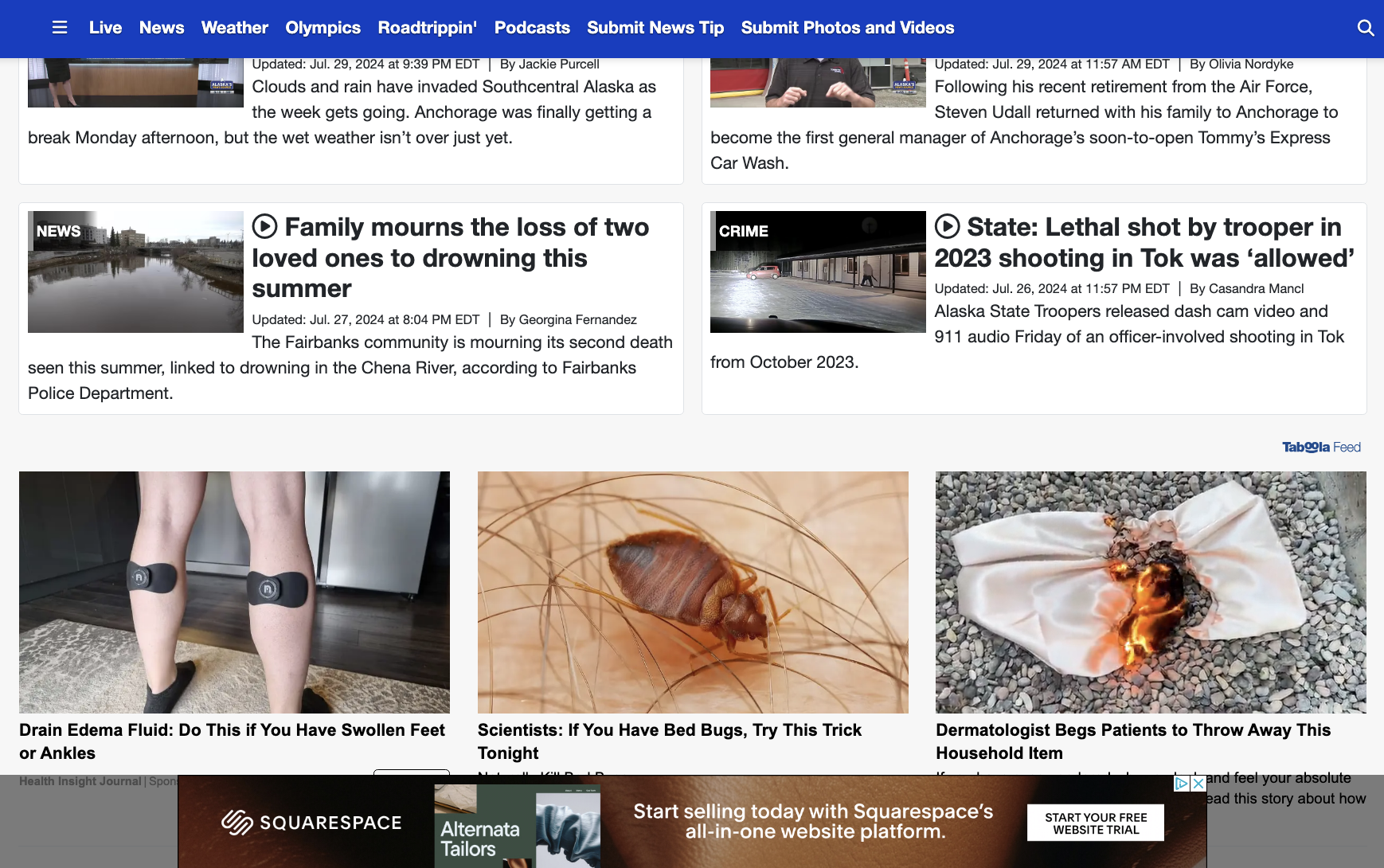}}
    \begin{tikzpicture}[overlay]
          \draw[red,thick,rounded corners] (0.1,0.8) rectangle (5.2,1.8);
          \draw[red,thick,rounded corners] (0.6,0.4) rectangle (4.8,0.7);
    \end{tikzpicture}
    \vspace{-.2in}
    \caption{Control}
    \label{fig:normal}
\end{subfigure}
\begin{subfigure}{.34\textwidth}
    \fbox{\includegraphics[width=.95\linewidth]{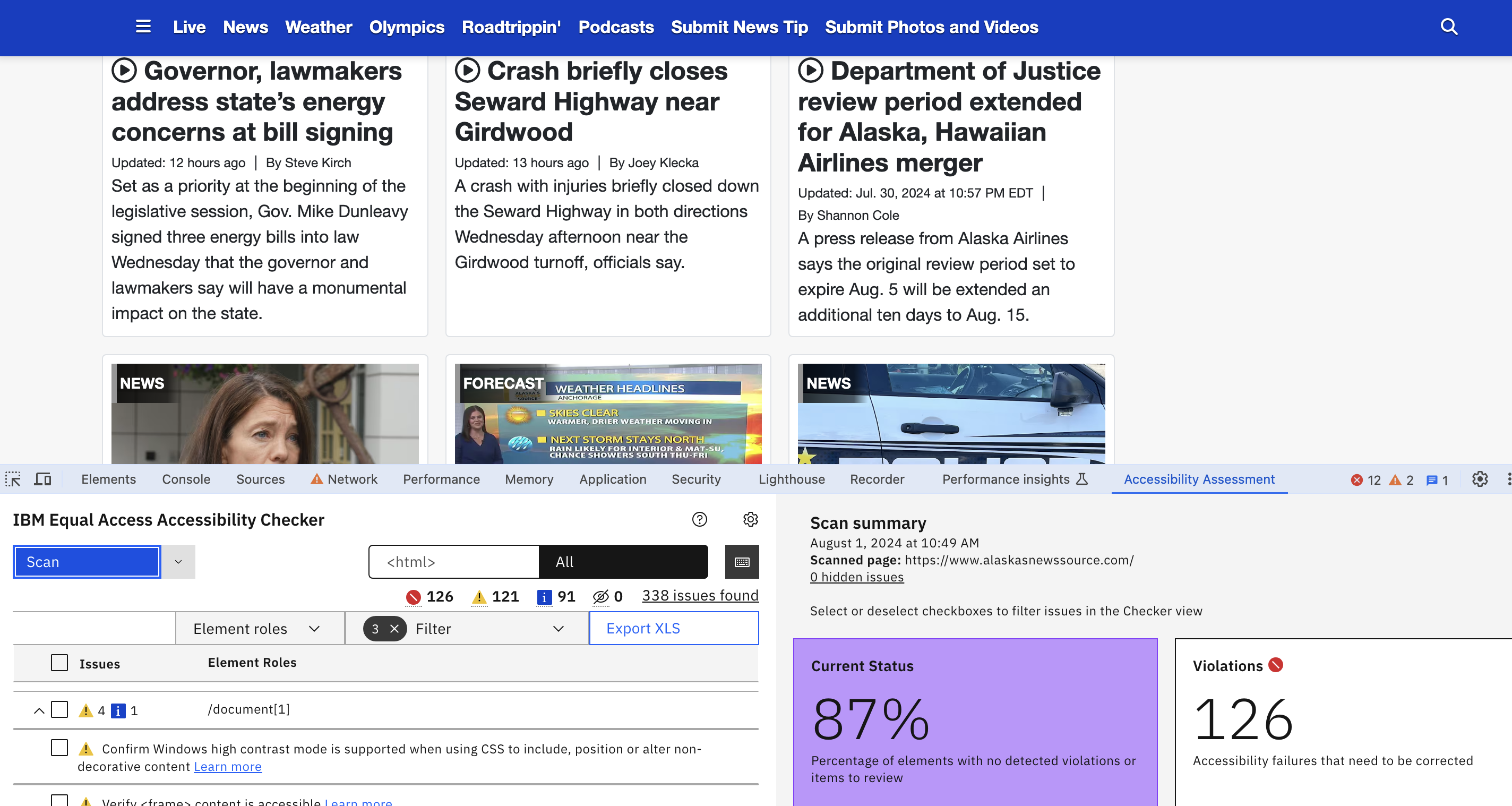}}
    \begin{tikzpicture}[overlay]
          \draw[red,thick,rounded corners] (4.7,0.5) rectangle (5.6,1);
    \end{tikzpicture}
    \vspace{-.18in}
    \caption{Accessibility without Ads}
\end{subfigure}
\begin{subfigure}{.34\textwidth}
    \fbox{\includegraphics[width=.96\linewidth]{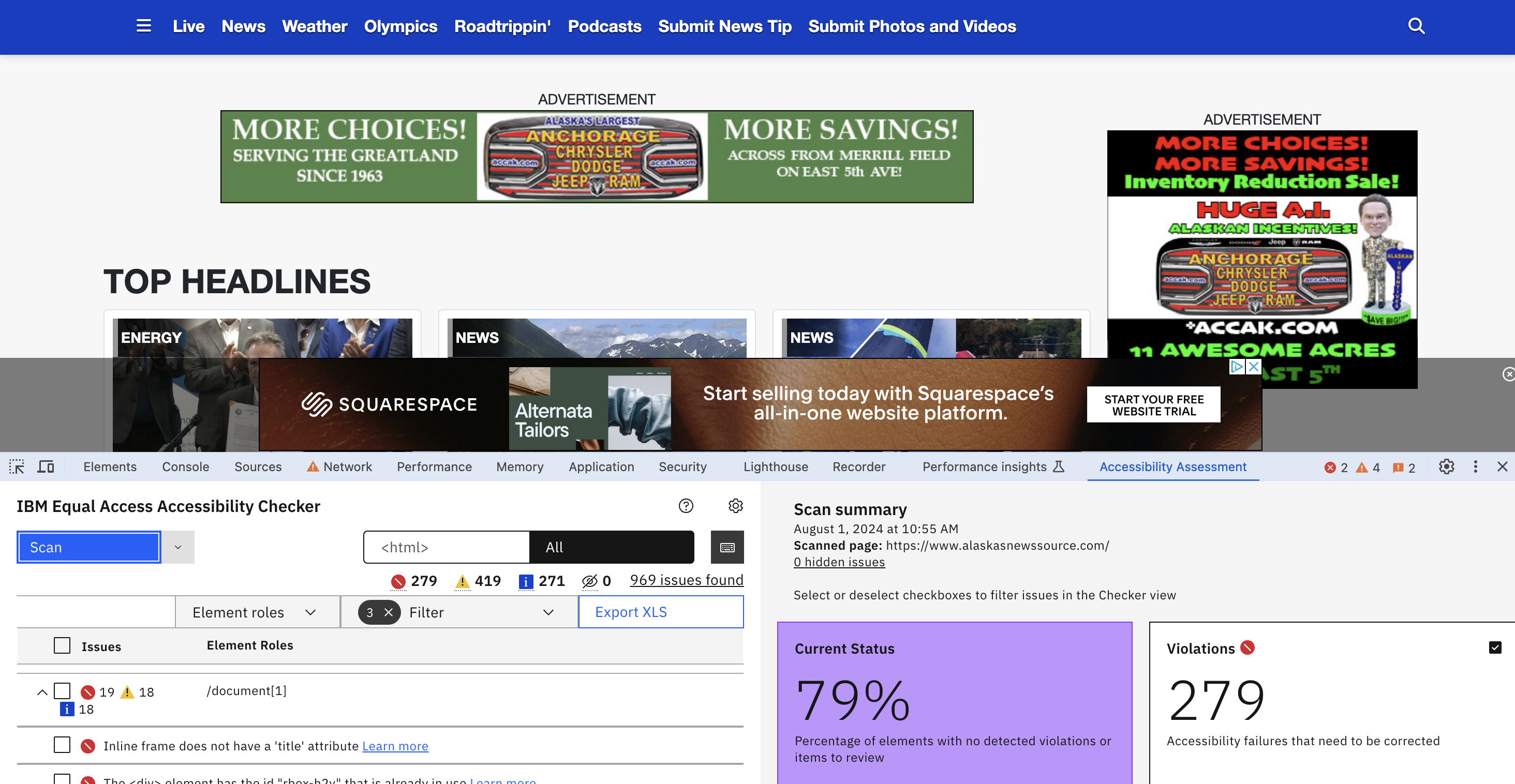}}
    \begin{tikzpicture}[overlay]
          \draw[red,thick,rounded corners] (4.6,0.5) rectangle (5.5,1);
    \end{tikzpicture}
    \vspace{-0.2in}
    \subcaption{Accessibility with Ads}
    \end{subfigure}

\caption{Accessibility report for {\tt alaskasnewssource.com}, highlighting the count of accessibility violations, which are 126 when the website is visited without ads and 279 when inspected with ads.} 
\vspace{-3ex}
\label{fig:motivating-example}
\end{figure*}

Approximately 15\% of the global population has some form of disability, and their access to the internet depends largely on website accessibility. Many of these users rely on assistive technologies like screen readers, which require highly accessible websites (\eg properly labeled images) to function effectively\cite{brown1992assistive}.
The Web Content Accessibility Guidelines (WCAG) \cite{WCAG-21}, led by the World Wide Web Consortium (W3C), have driven efforts to improve website's accessibility. As a result, numerous tools and services now exist to test and enhance website accessibility \cite{kumar2021comparing,gleason2020twitter,zhang2021screen, prakash2024all}. While website's accessibility is gradually improving due to these tools, the accessibility of website advertisements (ads) remains largely unexamined. 

Website ads are the driving force behind the free and open Internet\cite{evans2009online}. Website developers integrate ads on their websites by leasing space to ad-serving technologies (\eg Taboola \cite{Taboola}) that serve user-relevant ads when a visitor accesses the site. Consequently, developers give up control over the ad content served on their websites and whether it meets WCAG accessibility guidelines.

\noindent {\bf Problem.} Including ads on a website poses several challenges in evaluating the website's accessibility. 
First, popular website accessibility tools, like Google Lighthouse \cite{Google-lighthouse}, do not include ads in their assessments, which overestimates accessibility in evaluation reports. 
Second, due to the dynamic nature of ad content, website developers cannot predict which ads will be served to users or whether those ads will adhere to WCAG guidelines. Even when a website developer makes the best effort to ensure WCAG compliance, ads can significantly deteriorate the overall accessibility of the website.
Third, inaccessible website ads prevent users with disabilities from accessing a significant portion of the website. This is especially alarming when ads pose privacy and security risks \cite{bashir2016tracing, englehardt2016online, eslami2018communicating,li2012knowing, rastogi2016these}, as they are more likely to lead to fraud, phishing, and scams.
Fourth, blanket ad blocking harms the free internet and causes fundamental equity issues by making ads completely unreachable. Accessible ads not only encourage a free and open internet but also equip users with disabilities with the necessary context for safe and informed interaction with ads.

\noindent {\bf Prior-work.} While website accessibility has been extensively studied\cite{hackett2003accessibility, campoverde2023accessibility, mohammadi2024accessibility, agrawal2019evaluating, teixeira2021diversity, fok2022large, milne2014accessibility, ross2018examining}, to date, no comprehensive study has examined the accessibility of website ads at a large scale, leaving a significant gap in our understanding of website accessibility as a whole.
Only three studies have examined digital ad accessibility, focusing on small datasets of platform-specific ads and solely on blind website users. He et al.\cite{he2024tend} studied only 500 ad screens within Android apps and introduced a tool for detecting accessibility violations. Similarly, prior work manually examined {\tt alt text} on 67 websites \cite{thompson2001accessibility} and used a survey to understand screen reader users' challenges with website ads \cite{10.1145/3597503.3639229}. With the prevalence of website ads and their potential safety risks, the impact of website ads on accessibility remains an open yet important question.

\noindent {\bf Contributions.} In this paper, we conduct the first-of-its-kind large-scale, in-depth investigation of website ad accessibility and its impact on overall website accessibility. We perform an automated analysis of 430K website elements with nearly 100K website ad elements on the top 5K news and media websites. We first seek to understand the prevalence of inaccessible ads on the website and how they impact the overall accessibility of websites.
We build an automated pipeline to rigorously analyze the accessibility of website elements and the ads using IBM Equal Access Accessibility Checker and state-of-the-art ad-blocker, uBlockOrigin. Among ads that adversely impact a website's accessibility, we analyze the accessibility violations in accordance with WCAG guidelines. Identifying such violations provides key insights into commonly overlooked accessibility guidelines in ads and determines whether these violations are prevalent only in ads. To inform developers about accessibility violations across different ad-serving technologies, we further analyze inaccessible ads with respect to the technologies serving them.

Ads are historically more likely to lead to fraud, phishing, and scams~\cite{li2012knowing, rastogi2016these}. Even when ads are accessible, we investigate whether the provided accessible information truly represents the content and intent of the ad. We utilize open-source CLIP (Contrastive Language–Image Pretraining) model \cite{OpenAI-Clip} to confirm whether the content of ad images aligns with the provided alternative text for accessible ads. For instance, users with visual impairments solely rely on alternate text to access website content; false and misrepresented alternatives can mislead such users into involuntary interaction with ads. The onus for addressing inaccessible and misleading ads falls on the website developer whose website is hosting such ads.

%

\noindent {\bf Findings.}
Our analysis reveals that 75\% of websites have ads that frequently fail to comply with WCAG guidelines. On average, we observe  11 more accessibility violations due to ads. 
We further categorize the most common WCAG guideline violations found in ads to illustrate their potential impact on individuals with disabilities. We identified 28 different WCAG guideline violations in ad elements, with 7 of these violations predominantly occurring in only ad elements. These violations are related to missing captions, missing focus order, and missing page language, all of which are critical in ensuring smooth navigation on a website. 
Our analysis also identifies the ad-serving technologies responsible for the majority of these violations. We found that 52\% of websites use one of Google's ad-serving technologies (e.g., DoubleClick), which account for over 14K violations in their ad elements.

We show that the misrepresentation of data in alternate texts compared to the actual image deceives users with visual impairments and cognitive disabilities, making it difficult for them to interpret the intent of the ads. We found that 27\% of ad images misrepresent their content and intent. Furthermore, our manual inspection of ads from top-50 news websites reveals that up to 94\% of the ads collect trackable user identifiers upon {\tt onmouseover}, {\tt onhover}, and {\tt onclick}  events. These events help assistive technologies identify website elements that a user with disability desires to access. Thus, navigating ads can compromise the privacy of assistive technology users. To the best of our knowledge, this is the first work that investigates the accessibility violations of website ads on a large scale. 

\noindent {\bf Data Availability:} Our data is available at \url{https://zenodo.org/records/13137429}.

\begin{figure}[!t]
\begin{subfigure}{.23\textwidth}
    \fbox{\includegraphics[width=.99\linewidth, height=2.6cm]{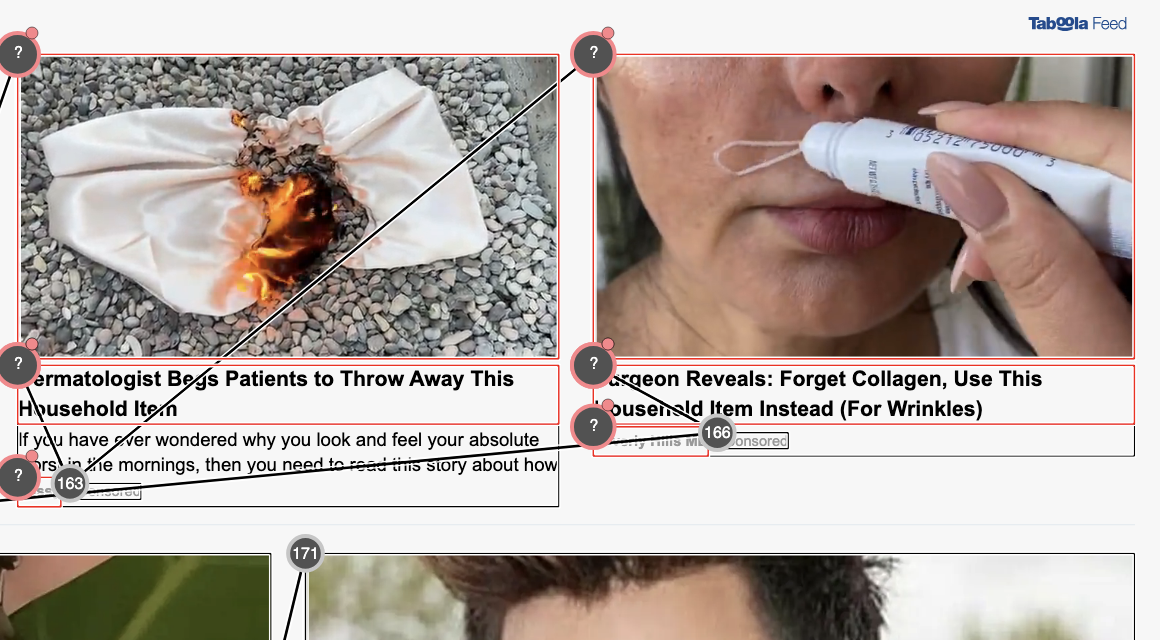}}
    \caption{Accessibility Violations in Ad Slots of Taboola.}
\end{subfigure}
\hspace{.05in}
\begin{subfigure}{.23\textwidth}
    \fbox{\includegraphics[width=.99\linewidth]{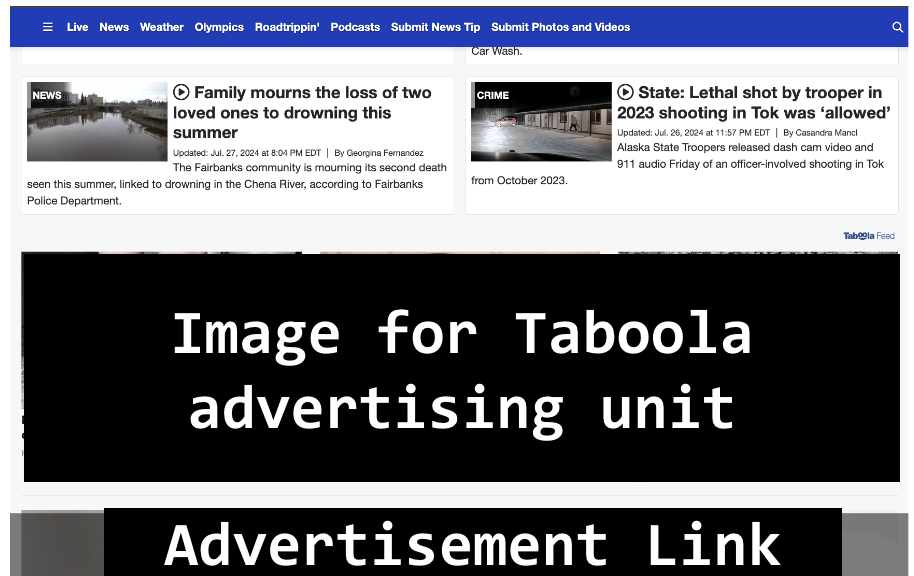}}
    \caption{Ads are not accessible to users relying on screen readers.}
\end{subfigure}
\caption{Violation identified in Taboola ad slots and its impact on how screen reader users experience the website.} 
\vspace{-4ex}
\label{fig:ad-issues-sample}
\end{figure} 
\begin{figure*}[h]
    \centering
      \includegraphics[width=0.9\textwidth]{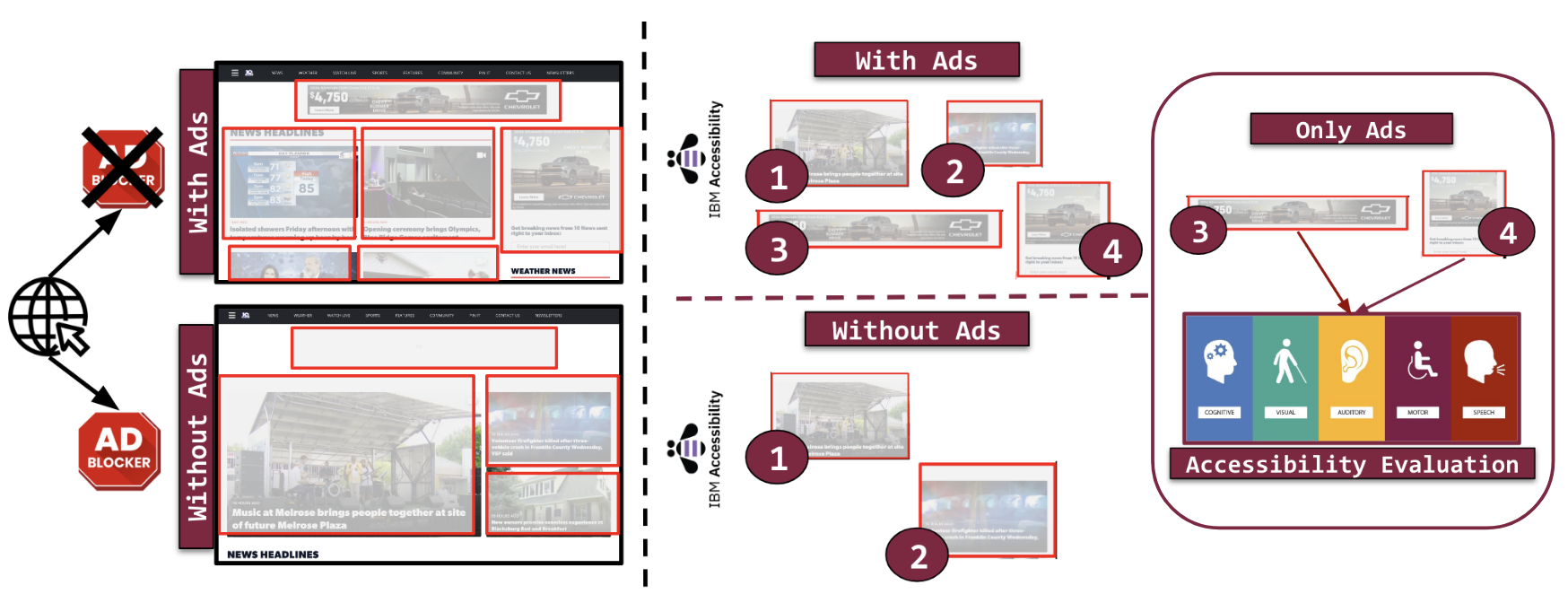}
        \caption{Overview of our methodology involving generating accessibility reports for each website with (\rwa) and without ads (\rwoa), then comparing them to isolate and analyze the accessibility violations caused by only ads ($R_{ads-only}$).}
        \label{fig:overview}
        \vspace{-.2in}
\end{figure*}
\section{Motivating Example}
\label{sec:motivation}
This section presents a case study showing how some ads on websites violate accessibility guidelines and affect the overall website's accessibility.
Alaska's News Source, {\tt alaskasnewssource.com}\cite{alaskanews}, provides local news, weather updates, sports coverage, and community events for the state of Alaska.
It is ranked 14\textsuperscript{th} in the top 5K in the news and media category.
In the stock Chrome browser, this website shows numerous ads in ad slots—three at the top and one at the bottom, as seen in Figure \ref{fig:motivating-example}(a). 
The top slots are served by Taboola\cite{Taboola} and the bottom by Google Ads\cite{Google-ads}.

\noindent\textbf{Website accessibility without ads.} 
In the first step, we generate the accessibility report for {\tt alaskasnewssource.com} by computing the violation count without including any ads.
We remove ads from the website to emulate the website development stage, where website developers are unaware of the precise ads that will be served by the ad-serving technologies. 
This is a common practice employed by many website quality and accessibility checkers like Google Lighthouse \cite{Google-lighthouse} that do not include ads in their accessibility evaluations. 
We use IBM's accessibility checker chrome extension, a tool used by over 20K developers\cite{IBM-accessibility}, mainly because it evaluates all website elements, including ad elements. 
To remove ads from the website, we deploy the uBlock Origin's Chrome extension, which has over 700K users\cite{uBlock-origin}, to block all ads on the website.  
As seen in Figure \ref{fig:motivating-example}(b), all ads on {\tt alaskasnewssource.com} are removed, and the accessibility checker reports 126 violations.
For example, the element in Listing \ref{lst:motivation} violates the WCAG 2.4.4 (Link Purpose) \cite{WCAG-list} guideline because it lacks descriptive link text, label, or an image with a text alternative. 
The guideline requires that links must have a clear purpose that users can understand, especially when the link is not accompanied by any other textual context. 
This can be resolved by adding {\tt aria-label="Download the Alaska's News Source Streaming Apps"} to this tag by developer. 
\lstdefinestyle{base}{
language=html,
moredelim=**[is][\color{red}]{@}{@},
moredelim=**[is][\color{darkgreen}]{~}{~},
numbers=left,
numberstyle=\tiny,
stepnumber=1,
numbersep=5pt,
escapechar=`
}
\begin{lstlisting}[caption={Anchor tag missing descriptive text},language=html, label={lst:motivation}, style=base]
<a data-tb-link="" target="\_self" rel="" href="https://www.alaskasnewssource.com/page/how-to-download-the-alaskas-news-source-streaming-apps/" class="link | text-reset d-block h-100 w-100">
\end{lstlisting}
Most accessibility violations can easily be addressed by the developer with simple refactoring of the website's code (\ie HTML, CSS, and JS). 

\noindent\textbf{Website accessibility with ads.} Next, we compute accessibility report for {\tt alaskasnewssource.com} by computing the violation count with ads, as displayed to the user.
This setting allows us to demonstrate the increase in the violations attributed to ads on the website.

Figure \ref{fig:motivating-example}(c) shows that ads appear on the website as they did on a normal visit. 
The accessibility checker reports 279 violations, showing a 121.4\% change in the violation count. 
Given that the only difference between the two websites is the presence of ads, it is reasonable to attribute the increase in accessibility violations to the ads.
We further examine specific violations related to ads, particularly concerning the ad slots Taboola serves. 

Figure \ref{fig:ad-issues-sample}(a) shows violations identified by the accessibility checker across three ad slots for Taboola (highlighted with a red circle around their title and short description).
This violates the WCAG 2.1.1 (Keyboard) guideline\cite{WCAG-list} because the element cannot be interacted with using a keyboard alone.
All ad images are designed with a {\tt link role}, which does not include a tabbable element. 
Thus, users who rely on keyboard navigation cannot directly access these elements. 
Additionally, most of these ad images do not include alternate text to make them accessible for visually impaired users.   
We further test the accessibility of these ads with  Google's Chrome Vox screen reader\cite{Chrome-vox}.
Vox only announces {\tt image for Taboola advertising unit} on tab focus without reading the title and short description of each ad. 
Figure \ref{fig:ad-issues-sample}(b) shows how the website and ads appear to visually impaired users using screen readers, leaving ads' content mostly inaccessible except the two vague alternate texts.

\textbf{\textit{It is difficult for a website developer to anticipate and address ad accessibility violations as these violations are non-deterministic and occur after the deployment of the website. }} 
In the case of {\tt alaskasnewssource.com}, ads are negatively impacting accessibility, which raises the concern of whether such an effect is prevalent across all ads: (1) if yes, then what kinds of violations do they cause? and (2) if not, then does the accessible information in these ads adequately represent their content? Similarly, it is useful to know which ad-serving technologies are serving more inaccessible ads.

\section{Research Questions}
Given the proliferation of ads on popular websites, their impact on vulnerable users, and the gap in developers' understanding of ad accessibility, we aim to answer the following research questions:
\begin{enumerate}
    \item \textbf{RQ1:} How do advertisements impact the accessibility guideline violations of websites?
    \item \textbf{RQ2:} What are the most common accessibility guideline violations found in advertisements?
    \item \textbf{RQ3:} Which ad-serving technologies frequently exhibit accessibility violations in their served advertisements?
    \item \textbf{RQ4:} How is information misrepresented in the alternate text of advertisements?
\end{enumerate}

\section{Methodology}
\label{sec:methodology}
To address the research questions, we designed an automated pipeline to inspect websites, identify accessibility violations in each element, extract violations specific to ad elements, and categorize them based on WCAG guidelines \cite{WCAG-21}.
%
Figure \ref{fig:overview} illustrates the overview of the process used in our empirical investigation, which is completely automated and packaged into a single measurement tool. 
\begin{table}[t]
\small
    \centering
    \scalebox{0.8}{\begin{tabular}{|>{\raggedright\arraybackslash}p{1.2cm}|>{\raggedright\arraybackslash}p{4cm}|>{\raggedright\arraybackslash}p{4.6cm}|}
\hline
{\bf Column} &  \textbf{Description} & \textbf{Example}\\ 
\hline
{\bf Page Title } &  Identifies the page or html file that was scanned. & Fox News\\ 
\hline
{\bf Page URL } & Identifies the page URL that was scanned.& www.foxnews.com\\ 
\hline
{\bf Element } & Type of HTML element where the issue is found.& {\tt <video>}\\ 
\hline
{\bf Element Code} & Code of HTML element where the issue is found.& {\tt <video hola-pid="4" loop="" muted="" autoplay="" playsinline="">}\\
\hline
{\bf Xpath } & Xpath of the HTML element where the issue is found. & {\tt /html[1]/.../.../video[1}]\\ 
\hline
{\bf Check point } & WCAG check points this issue falls into. & 2.1.1 Keyboard\\
\hline
{\bf WCAG level } & A, AA or AAA. WCAG level for this issue.& A\\ 
\hline
\end{tabular}}
    \caption{Sample accessibility evaluation report detailing elements and associated violation information.}
    \label{table:report}
    \vspace{-0.25in}
\end{table}
\subsection{Websites Visited}
We visit the top 5K news and media category websites from SimilarWeb \cite{Similar-web}, using a custom Selenium (version 4.11.2) based automated website navigator running on Google Chrome (version 120.0.6099.129).  
Prior literature has also focused on this category of websites because they have the most ads \cite{thompson2001accessibility,zeng2020bad,he2024tend}, offering the most breadth and coverage of ads. 
For every website, we launch our automated website navigator to first visit the website and then eventually trigger the accessibility evaluation using the IBM accessibility tool, as explained in Section~\ref{sec:accessiblity-evaluation}. 
It takes approximately 30 seconds for a website to fully load (until the onLoad event is fired), on average, and an additional five minutes to ensure all ads are loaded before applying the accessibility checks.  
We use a stateless crawling process, with all cookies and local browser states cleared between consecutive visits. 
This approach helps ensure that the collected accessibility data accurately reflects the current state of the website without being biased by previous visits. 
We visited all 5K websites in North America. 

We crawl each website twice to obtain two accessibility evaluation reports for the same website:
\begin{enumerate}
    \item $Report (R_{with-ads})$: Accessibility report of the website with ads.
    \item $Report (R_{without-ads})$: Accessibility report of the website without ads.
\end{enumerate}
$Report (R_{with-ads})$ contains the accessibility report for all elements on the website, including ad elements.
$Report (R_{without-ads})$ contains the accessibility report for all elements on the website, excluding ad elements.
Finally, we extract the $ Report (R_{only-ads})$ by differencing the 
$R_{with-ads}$ and $R_{without-ads}$, which represents the accessibility report for the ad elements only:
\begin{equation}
    \underbrace{R_{with-ads}}_{\text{Website elements with Ads}} - \underbrace{R_{without-ads}}_{\text{Website elements without Ads}} = \underbrace{R_{only-ads}}_{\text{Ad elements}}
\end{equation}
\noindent\textbf{Why do we need this approach to find the accessibility of ads?} To date, there does not exist any tool that can identify and isolate accessibility scores for ad elements alone. 
Ads can only be evaluated post-deployment because, during the development phase, ad-serving technology only requires reserved space on the website. 
The actual ads only appear on the website based on user client profiling and real-time advertiser bidding. 
We opt to evaluate these ads post-deployment on the client side to understand the true accessibility impact of ads, offering insights for developers and ad-serving technologies. 

\begin{table}[t]
\small
    \centering
    \scalebox{0.85}{\begin{tabular}{|>{\raggedright\arraybackslash}p{1.2cm}|>{\raggedright\arraybackslash}p{1cm}|>{\raggedright\arraybackslash}p{4.9cm}|>{\raggedleft\arraybackslash}p{1cm}|}
     \cline{1-4}
     {\bf Violation } & {\bf Element} & {\bf WCAG} & {\bf Count}\\ 
     \cline{1-4}
        \multirow{2}{*}{Overall} & All & - & 430,497 \\
                  & Ad & - & 99,422 \\
        \cline{1-4}
        \multirow{3}{*}{Levels} & \multirow{3}{*}{Ad} & A  & 80,220 \\
                &  & AA  & 43,591 \\
                
        \hline
        \multirow{30}{*}{Criterion} & \multirow{30}{*}{Ad} & 2.4.7 Focus Visible& 41,110 \\ &  & 2.4.1 Bypass Blocks& 37,357 \\ 

                &  & 3.2.2 On Input & 26,629 \\
                & & 1.1.1 Non-text Content& 16,387 \\
                &  & 4.1.2 Name, Role, Value& 15,712 \\
                &  & 2.4.4 Link Purpose (In Context)& 7,714 \\
                &  & 1.3.4 Orientation& 6,493 \\
                &  & 1.3.1 Info and Relationships& 6,118 \\
                &  & 2.1.1 Keyboard& 6,026 \\
                &  & 2.4.3 Focus Order& 3,575 \\
                &  & 1.4.3 Contrast (Minimum)& 2,890 \\
                &  & 2.5.3 Label in Name& 1,190 \\
                &  & 1.2.1 Audio-only and Video-only& 1,147 \\
                &  & 1.2.4 Captions (Live)& 1,144 \\
                &  & 1.4.1 Use of Color& 1,021 \\
                &  & 3.3.2 Labels or Instructions& 269 \\
                &  & 1.3.3 Sensory Characteristics& 92 \\
                &  & 3.1.1 Language of Page& 60 \\
                &  & 1.4.12 Text Spacing & 26 \\
                &  & 1.4.4 Resize text& 21 \\
                &  & 2.1.2 No Keyboard Trap& 17 \\
                &  & 3.2.1 On Focus& 13 \\
                &  & 2.4.2 Page Titled& 6 \\
                &  & 2.4.6 Headings and Labels& 6 \\

                &  & 1.3.2 Meaningful Sequence& 3 \\

                &  & 1.3.5 Identify Input Purpose& 2 \\
                &  & 2.2.1 Timing Adjustable & 2\\
                &  & 1.4.13 Content on Hover& 1 \\

     \cline{1-4}
\end{tabular}}
    \caption{Number of website elements with violations during the campaign, detailing ad elements by violation level and criteria.}
    \label{table:dataset}
    \vspace{-0.25in}
\end{table}
\subsection{Blocking Website Advertisements}
To measure the website's accessibility without ads, we aim to load the website without any ads in our crawler, which requires removing all ads before performing accessibility evaluations. 
We leverage a widely used, advanced ad-blocking tool called uBlock Origin\cite{uBlock-origin} to remove ads from a website before it completely loads in the browser. 
uBlock Origin relies on a crowd-sourced filter list\cite{easylist, easyprivacy}, which are databases of rules and patterns that identify and block resources from commonly used ad-serving technologies. 
Almost all ad blocking tools\cite{uBlock-origin, Adblockplus, adguard} rely on the same filter lists and, therefore, have more or less similar effectiveness in blocking ads.  
Internally, when a user visits a website, the ad blocker scans the page and its resources, comparing URLs and scripts against its filter lists.
If it detects elements associated with ad-serving technology, it prevents these elements from being loaded on the page. 
This can involve blocking requests to known ad servers, hiding ad elements within the DOM, or preventing ad scripts from executing. 
As a result, the ad content is not displayed to the user, resulting in an ad-free website, which we use to perform accessibility evaluations, $R_{without-ads}$. 
Prior work\cite{castell2022demystifying} on the effectiveness of ad-blocking tools has shown that uBlock Origin is the most effective ad blocker. 

\subsection{Accessibility Evaluation}
\label{sec:accessiblity-evaluation}
After the website loads, we automatically perform a rigorous accessibility evaluation using the IBM Equal Access Accessibility Checker \cite{IBM-accessibility}, a Chrome extension. 
Initially, we use Google's Lighthouse\cite{Google-lighthouse}, commonly used for quality checks during development. 
However, tools like Lighthouse, Axe \cite{Axe}, and Wave \cite{Wave} exclude ad elements by default, leading to inaccurate measurements for sites with ads. 
Thus, we use the IBM checker to assess accessibility for all website elements, including ads.
Internally, IBM's accessibility checker injects a content script into the website to interact with the website's DOM. 
The content script scans the website's DOM to identify elements and runs a series of checks based on the following predefined rules such as WCAG 2.1 \cite{WCAG-21}.
The results are intercepted by our automated website navigator and stored in separate storage for downstream analysis. 
\noindent\textbf{Accessibility Report.} 
Table \ref{table:report} shows a sample accessibility evaluation report with a detailed analysis of violations, including page title, URL, HTML element type, element code, XPath, WCAG checkpoint, and violation level. Each checkpoint corresponds to a specific WCAG criterion with a description of the violation.
%
The WCAG level indicates the priority and severity of accessibility violations, categorized into three levels:
\begin{itemize}
    \item \textbf{Level A:} These guidelines are fundamental for all websites. Non-compliance makes website content inaccessible to users with disabilities. For example, text alternatives for non-text content (WCAG 1.1.1) are crucial for users with visual impairments. Meeting Level A criteria is a top priority. 
    \item \textbf{Level AA:} These guidelines should be met by most websites. Conformance to Level AA significantly improves the accessibility of website content. For example, ensuring that text and images of text have a contrast ratio of at least 4.5:1 (WCAG 1.4.3) is crucial for users with visual impairments, including those with color blindness.
    \item \textbf{Level AAA:} These guidelines represent the highest website accessibility standards for full accessibility. For example, sign language interpretation for pre-recorded audio (WCAG 1.2.6) benefits deaf users. Level AAA conformance is not mandatory for WCAG compliance as it applies to specific content types. \textit{Note: These violations require manual assessment and are excluded from this analysis}.
    
\end{itemize}
%


\noindent\textbf{Extracting Ad-elements.} We first generate the two accessibility reports of each website in our dataset, $R_{with-ads}$ and $R_{without-ads}$.
$R_{with-ads}$ represents the accessibility report of all the website elements, including ads, as illustrated in Figure \ref{fig:overview} \ding{202} - \ding{205}.
$R_{without-ads}$ represents the accessibility report of all the website elements excluding ads, as illustrated in Figure \ref{fig:overview} \ding{202} - \ding{203}.
Finally, we automatically compare the two accessibility reports by first matching website elements based on their code and Xpath location, resulting in a list of website elements that are not ads (\ding{202} and \ding{203}). 
We exclude these elements from $R_{with-ads}$, which results in an accessibility report of ads only, $R_{ads-only}$ as illustrated in Figure \ref{fig:overview} \ding{204} - \ding{205}.
%
We analyzed 430K unique HTML elements with accessibility violations across 5k websites, of which 23.1\% are ad elements.
Computing accessibility reports is time-consuming and resource-intensive due to the extensive checks required on the rendered page, styling, and HTML structure.
%
It also cannot be deployed on the cloud, as certain accessibility metrics can only be measured when active display is available for UI rendering.
Therefore, analysis on the 5K website consumed more than 484 CPU hours to generate accessibility reports spanning 20.6 GB.  
Table \ref{table:dataset} lists the number of website elements with violations that we interacted with during this campaign.
Specifically, it details the ad elements with violations, categorizing them by the level and criteria of violations.

\section{Results}
\label{sec:results}
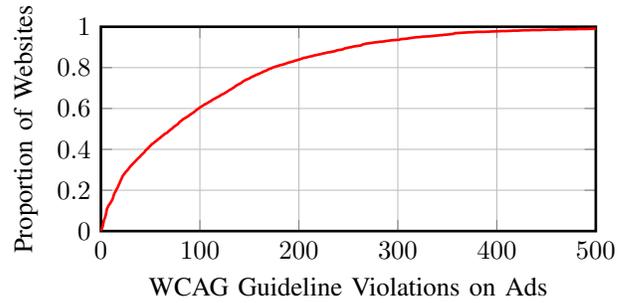
\begin{figure}
    \centering
    \begin{tikzpicture}
\begin{axis}[
    xlabel={WCAG Guideline Violations on Ads},
    ylabel={Proportion of Websites},
    legend pos=north west,
    ymin=0,
    ymax=1,
    xmin=0,
    xmax=500,
    grid=both,
    line width=1pt,
    width=0.45*\textwidth,
    height=4.3cm,
]

\addplot[color=red, solid] coordinates {
(0.000, 0.002)
(1.000, 0.011)
(2.000, 0.027)
(3.000, 0.052)
(4.000, 0.064)
(5.000, 0.080)
(6.000, 0.107)
(7.000, 0.118)
(8.000, 0.127)
(9.000, 0.133)
(10.000, 0.141)
(11.000, 0.149)
(12.000, 0.159)
(13.000, 0.176)
(14.000, 0.188)
(15.000, 0.196)
(16.000, 0.206)
(17.000, 0.214)
(18.000, 0.226)
(19.000, 0.236)
(20.000, 0.248)
(21.000, 0.258)
(22.000, 0.269)
(23.000, 0.276)
(24.000, 0.282)
(25.000, 0.289)
(26.000, 0.294)
(27.000, 0.299)
(28.000, 0.305)
(29.000, 0.312)
(30.000, 0.318)
(31.000, 0.323)
(32.000, 0.328)
(33.000, 0.333)
(34.000, 0.338)
(35.000, 0.343)
(36.000, 0.348)
(37.000, 0.353)
(38.000, 0.358)
(39.000, 0.362)
(40.000, 0.367)
(41.000, 0.373)
(42.000, 0.377)
(43.000, 0.382)
(44.000, 0.386)
(45.000, 0.394)
(46.000, 0.398)
(47.000, 0.403)
(48.000, 0.407)
(49.000, 0.413)
(50.000, 0.417)
(51.000, 0.423)
(52.000, 0.427)
(53.000, 0.430)
(54.000, 0.434)
(55.000, 0.439)
(56.000, 0.441)
(57.000, 0.446)
(58.000, 0.450)
(59.000, 0.453)
(60.000, 0.457)
(61.000, 0.462)
(62.000, 0.465)
(63.000, 0.469)
(64.000, 0.474)
(65.000, 0.477)
(66.000, 0.479)
(67.000, 0.482)
(68.000, 0.487)
(69.000, 0.491)
(70.000, 0.495)
(71.000, 0.499)
(72.000, 0.502)
(73.000, 0.507)
(74.000, 0.512)
(75.000, 0.514)
(76.000, 0.518)
(77.000, 0.521)
(78.000, 0.525)
(79.000, 0.532)
(80.000, 0.536)
(81.000, 0.539)
(82.000, 0.543)
(83.000, 0.548)
(84.000, 0.550)
(85.000, 0.553)
(86.000, 0.556)
(87.000, 0.559)
(88.000, 0.562)
(89.000, 0.565)
(90.000, 0.570)
(91.000, 0.572)
(92.000, 0.576)
(93.000, 0.579)
(94.000, 0.582)
(95.000, 0.586)
(96.000, 0.591)
(97.000, 0.594)
(98.000, 0.598)
(99.000, 0.601)
(100.000, 0.605)
(101.000, 0.608)
(102.000, 0.611)
(103.000, 0.614)
(104.000, 0.617)
(105.000, 0.621)
(106.000, 0.622)
(107.000, 0.625)
(108.000, 0.628)
(109.000, 0.631)
(110.000, 0.634)
(111.000, 0.636)
(112.000, 0.640)
(113.000, 0.643)
(114.000, 0.646)
(115.000, 0.648)
(116.000, 0.651)
(117.000, 0.654)
(118.000, 0.656)
(119.000, 0.659)
(120.000, 0.661)
(121.000, 0.665)
(122.000, 0.668)
(123.000, 0.670)
(124.000, 0.672)
(125.000, 0.675)
(126.000, 0.678)
(127.000, 0.681)
(128.000, 0.684)
(129.000, 0.687)
(130.000, 0.691)
(131.000, 0.693)
(132.000, 0.696)
(133.000, 0.699)
(134.000, 0.702)
(135.000, 0.705)
(136.000, 0.710)
(137.000, 0.713)
(138.000, 0.715)
(139.000, 0.718)
(140.000, 0.721)
(141.000, 0.723)
(142.000, 0.726)
(143.000, 0.730)
(144.000, 0.734)
(145.000, 0.736)
(146.000, 0.739)
(147.000, 0.741)
(148.000, 0.744)
(149.000, 0.745)
(150.000, 0.747)
(151.000, 0.751)
(152.000, 0.753)
(153.000, 0.756)
(154.000, 0.757)
(155.000, 0.760)
(156.000, 0.762)
(157.000, 0.765)
(158.000, 0.766)
(159.000, 0.769)
(160.000, 0.770)
(161.000, 0.773)
(162.000, 0.775)
(163.000, 0.777)
(164.000, 0.780)
(165.000, 0.782)
(166.000, 0.784)
(167.000, 0.785)
(168.000, 0.788)
(169.000, 0.791)
(170.000, 0.793)
(171.000, 0.795)
(172.000, 0.796)
(173.000, 0.799)
(174.000, 0.801)
(175.000, 0.803)
(176.000, 0.804)
(177.000, 0.806)
(178.000, 0.807)
(179.000, 0.809)
(180.000, 0.810)
(181.000, 0.812)
(182.000, 0.813)
(183.000, 0.814)
(184.000, 0.815)
(185.000, 0.817)
(186.000, 0.818)
(187.000, 0.820)
(188.000, 0.822)
(189.000, 0.824)
(190.000, 0.825)
(191.000, 0.826)
(192.000, 0.828)
(193.000, 0.829)
(194.000, 0.830)
(195.000, 0.832)
(196.000, 0.834)
(197.000, 0.835)
(198.000, 0.836)
(199.000, 0.838)
(200.000, 0.839)
(201.000, 0.841)
(202.000, 0.842)
(203.000, 0.844)
(204.000, 0.846)
(205.000, 0.848)
(206.000, 0.848)
(207.000, 0.849)
(208.000, 0.850)
(209.000, 0.852)
(210.000, 0.853)
(211.000, 0.854)
(212.000, 0.855)
(213.000, 0.856)
(214.000, 0.857)
(215.000, 0.859)
(216.000, 0.860)
(217.000, 0.861)
(218.000, 0.862)
(219.000, 0.863)
(220.000, 0.864)
(221.000, 0.865)
(222.000, 0.866)
(223.000, 0.867)
(224.000, 0.869)
(225.000, 0.870)
(226.000, 0.870)
(227.000, 0.872)
(228.000, 0.873)
(229.000, 0.874)
(230.000, 0.875)
(231.000, 0.875)
(232.000, 0.876)
(233.000, 0.878)
(234.000, 0.878)
(235.000, 0.879)
(236.000, 0.881)
(237.000, 0.881)
(238.000, 0.884)
(239.000, 0.885)
(240.000, 0.886)
(241.000, 0.887)
(242.000, 0.887)
(243.000, 0.888)
(244.000, 0.889)
(245.000, 0.891)
(246.000, 0.893)
(247.000, 0.895)
(248.000, 0.896)
(249.000, 0.896)
(250.000, 0.898)
(251.000, 0.899)
(252.000, 0.900)
(253.000, 0.900)
(254.000, 0.902)
(255.000, 0.903)
(256.000, 0.904)
(257.000, 0.905)
(258.000, 0.905)
(259.000, 0.906)
(260.000, 0.907)
(261.000, 0.907)
(262.000, 0.908)
(263.000, 0.909)
(264.000, 0.912)
(265.000, 0.914)
(266.000, 0.915)
(267.000, 0.916)
(268.000, 0.918)
(269.000, 0.918)
(270.000, 0.919)
(271.000, 0.919)
(272.000, 0.920)
(273.000, 0.921)
(274.000, 0.921)
(275.000, 0.922)
(277.000, 0.924)
(278.000, 0.924)
(279.000, 0.925)
(280.000, 0.925)
(281.000, 0.926)
(282.000, 0.927)
(283.000, 0.927)
(284.000, 0.927)
(285.000, 0.928)
(286.000, 0.928)
(287.000, 0.929)
(288.000, 0.930)
(289.000, 0.931)
(290.000, 0.931)
(291.000, 0.932)
(292.000, 0.933)
(293.000, 0.933)
(294.000, 0.934)
(295.000, 0.934)
(296.000, 0.935)
(297.000, 0.935)
(298.000, 0.935)
(299.000, 0.936)
(300.000, 0.936)
(301.000, 0.936)
(302.000, 0.938)
(303.000, 0.939)
(304.000, 0.940)
(305.000, 0.940)
(307.000, 0.941)
(308.000, 0.941)
(309.000, 0.942)
(310.000, 0.942)
(311.000, 0.944)
(312.000, 0.944)
(313.000, 0.945)
(314.000, 0.946)
(315.000, 0.946)
(316.000, 0.947)
(317.000, 0.948)
(318.000, 0.948)
(319.000, 0.949)
(321.000, 0.949)
(322.000, 0.950)
(323.000, 0.950)
(324.000, 0.951)
(326.000, 0.952)
(327.000, 0.952)
(328.000, 0.952)
(329.000, 0.953)
(331.000, 0.953)
(332.000, 0.953)
(333.000, 0.954)
(334.000, 0.954)
(335.000, 0.955)
(336.000, 0.955)
(337.000, 0.956)
(338.000, 0.956)
(339.000, 0.957)
(340.000, 0.957)
(341.000, 0.958)
(342.000, 0.958)
(343.000, 0.959)
(345.000, 0.959)
(346.000, 0.960)
(347.000, 0.961)
(348.000, 0.961)
(349.000, 0.961)
(350.000, 0.962)
(351.000, 0.962)
(352.000, 0.963)
(354.000, 0.964)
(355.000, 0.965)
(356.000, 0.965)
(357.000, 0.967)
(358.000, 0.969)
(359.000, 0.969)
(360.000, 0.969)
(362.000, 0.970)
(363.000, 0.970)
(364.000, 0.970)
(365.000, 0.971)
(366.000, 0.971)
(369.000, 0.972)
(370.000, 0.972)
(371.000, 0.973)
(372.000, 0.973)
(373.000, 0.973)
(375.000, 0.973)
(377.000, 0.974)
(378.000, 0.974)
(379.000, 0.975)
(385.000, 0.975)
(386.000, 0.975)
(387.000, 0.975)
(388.000, 0.975)
(390.000, 0.976)
(392.000, 0.976)
(393.000, 0.976)
(394.000, 0.977)
(396.000, 0.977)
(397.000, 0.977)
(400.000, 0.978)
(402.000, 0.978)
(405.000, 0.979)
(406.000, 0.979)
(408.000, 0.979)
(409.000, 0.979)
(410.000, 0.979)
(413.000, 0.980)
(414.000, 0.980)
(415.000, 0.980)
(416.000, 0.981)
(417.000, 0.981)
(421.000, 0.981)
(422.000, 0.981)
(423.000, 0.982)
(424.000, 0.982)
(425.000, 0.982)
(427.000, 0.983)
(428.000, 0.983)
(429.000, 0.983)
(433.000, 0.983)
(435.000, 0.984)
(440.000, 0.984)
(442.000, 0.984)
(445.000, 0.984)
(446.000, 0.985)
(449.000, 0.985)
(450.000, 0.985)
(451.000, 0.986)
(459.000, 0.986)
(461.000, 0.986)
(464.000, 0.987)
(465.000, 0.987)
(466.000, 0.987)
(468.000, 0.988)
(470.000, 0.988)
(471.000, 0.988)
(474.000, 0.988)
(480.000, 0.988)
(482.000, 0.989)
(484.000, 0.989)
(488.000, 0.989)
(489.000, 0.989)
(490.000, 0.989)
(497.000, 0.990)
(501.000, 0.990)
(510.000, 0.990)
(511.000, 0.990)
(512.000, 0.990)
(514.000, 0.991)
(515.000, 0.991)
(519.000, 0.991)
(524.000, 0.991)
(526.000, 0.991)
(531.000, 0.992)
(532.000, 0.992)
(542.000, 0.992)
(545.000, 0.992)
(548.000, 0.993)
(549.000, 0.993)
(550.000, 0.993)
(560.000, 0.993)
(572.000, 0.993)
(581.000, 0.994)
(589.000, 0.994)
(592.000, 0.994)
(597.000, 0.994)
(605.000, 0.994)
(621.000, 0.995)
(630.000, 0.995)
(632.000, 0.995)
(634.000, 0.995)
(645.000, 0.995)
(658.000, 0.996)
(659.000, 0.996)
(664.000, 0.996)
(681.000, 0.996)
(686.000, 0.996)
(689.000, 0.997)
(753.000, 0.997)
(766.000, 0.997)
(781.000, 0.997)
(794.000, 0.998)
(838.000, 0.998)
(868.000, 0.998)
(884.000, 0.998)
(905.000, 0.998)
(978.000, 0.999)
(1022.000, 0.999)
(1062.000, 0.999)
(1078.000, 0.999)
(1159.000, 0.999)
(1260.000, 1.000)
(1310.000, 1.000)
(1840.000, 1.000)
};

\end{axis}
\end{tikzpicture}
    \vspace{-0.4in}
    \caption{CDF plot of websites by accessibility guideline violations in ad elements (\roa).}
    \label{fig:cdf-scores}
    \vspace{-0.2in}
\end{figure}
\begin{figure}
    \centering
    \begin{tikzpicture}
\begin{axis}[
    xlabel={WCAG Guideline Violations on Ads},
    ylabel={\# of Webistes},
    symbolic x coords={ 0, 10, 20, 30, 40, 50, 60, 70, 80, 90, 100, 110, 120, 130, 140, 150, 160, 170, 180, 190, 200, 210, 220, 230, 240, 250, 260, 270, 280, 290, 300, 310, 320, 330, 340, 350, 360, 370, 380, 390, 400, 410, 420, 430, 440, 450, 460, 470, 480, 490, 500, 510, 520, 530 },
    xmin = 10,
    ymin = 0,
    xmax = 400,
    width=1*\linewidth,
    height=4.5cm,
    grid=major,
    yticklabel style={rotate=90},
    ybar=1*\pgflinewidth,
   bar width=.08cm,
    ]
    \addplot[style={fill=red,mark=none}] coordinates {(0,2321) (10,555) (20,386) (30,230) (40,173) (50,162) (60,122) (70,113) (80,68) (90,68) (100,54) (110,52) (120,51) (130,42) (140,50) (150,32) (160,30) (170,43) (180,47) (190,70) (200,56) (210,36) (220,29) (230,19) (240,12) (250,20) (260,22) (270,6) (280,12) (290,6) (300,9) (310,5) (320,6) (330,4) (340,3) (350,2) (360,3) (370,1) (380,3) (390,3) (400,1) (410,2) (420,0) (430,1) (440,2) (450,2) (460,3) (470,0) (480,2) (490,1) (500,0) (510,2) (520,0) (530,0)};
\end{axis}
\end{tikzpicture}
    \vspace{-0.4in}
    \caption{Histogram of the number of websites by accessibility guideline violations in ad elements (\roa).}
    \label{fig:histogram-violations}
    \vspace{-0.2in}
\end{figure}
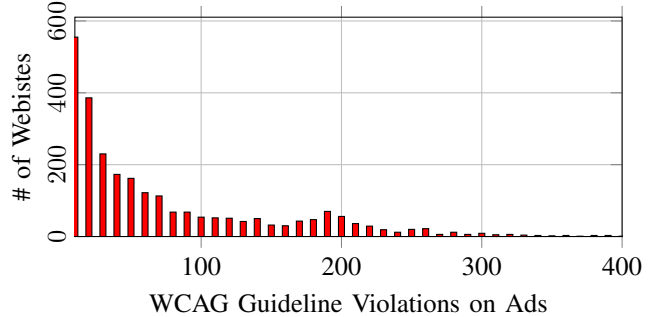
\subsection{RQ1: Impact of Advertisements on Accessibility Violations}
We first examine the impact of ads on the number of accessibility violations on websites.
This analysis demonstrates how third-party ad content can lower a website's accessibility, which has several downstream consequences, such as potentially low search rankings on Google due to accessibility violations\cite{Seo-accessibility} in addition to the fundamental inclusivity issues for users with disabilities. 
Figure \ref{fig:cdf-scores} presents the cumulative distribution of websites and their WCAG guideline violations of ad elements (\roa). 
%
Concretely, the average violation count is 106.2, with a median of 72, for websites without ads (\rwoa). 
For websites with ads (\rwa), the average violation count increases to 118.1, with a median of 81.
%
Overall, the average violation count for ad elements (\roa) is to 11.7. 
While the magnitude of the increase in violations may not look drastic, it is important to understand that a single violation directly points to an accessibility issue of a website element that makes it inaccessible for at least one type of disabled website user. 
As mentioned before, WCAG guidelines violations in ads are more challenging to address than the violations in original website content, as a developer can immediately act on violations from original website content, whereas, in the case of ads, it is infeasible for the developer to address the violation due to user-specific dynamic nature of ads. 

To further understand the number of violations caused by the ads on websites, we distribute websites with the number of violations caused by ads. 
Figure \ref{fig:histogram-violations} displays the distribution of websites for one or more accessibility guideline violations.  
We find that nearly 74.6\% of websites have at least one accessibility violation in their ad elements (\roa), and 66.7\% of websites have an increase in violations after including ads (\rwoa $\rightarrow$ \rwa).
\begin{tcolorbox}[colback=blue!5!white]
 {
    {\bf \em Finding.} Overall, nearly 75\% of websites show at least one accessibility violation attributed to ad elements, underscoring the need for further investigation into these violations.
 }
\end{tcolorbox}

\begin{table}[t]
\small
    \centering
    \scalebox{0.97}{\begin{tabular}{|>{\raggedright\arraybackslash}p{6.8cm}|>{\raggedright\arraybackslash}p{1.2cm}|}
\hline
\textbf{Violation in Ad elements (\roa) } & \textbf{Websites} \\
\hline
Check the keyboard focus indicator is visible when using CSS declaration for 'border' or 'outline' & 2,708 \\
\hline
Inform the user when their input action will open a new window or otherwise change their context & 2,358 \\
\hline
 Hyperlink has no link text, label or image with a text alternative & 1,915 \\
\hline
 Verify \texttt{<frame>} content is accessible & 1,808 \\
\hline
Confirm the element should be tabbable and if so, it becomes visible when it has keyboard focus & 1,632 \\
\hline
Content is not within a landmark element & 1,605 \\
\hline
Verify there is a way to bypass blocks of content that are repeated on multiple Web pages & 1,490 \\
\hline
Confirm Windows high contrast mode is supported when using CSS to include, position or alter non-decorative content & 1,427 \\
\hline
Page detected as HTML, but does not have a 'lang' attribute & 1,374 \\
\hline
Inline frame does not have a 'title' attribute & 1,310 \\
\hline
\end{tabular}

}
    \caption{Top-10 Most prevalent violations in ad elements}
    \label{table:rq2-violations}
    \vspace{-0.25in}
\end{table}
\begin{figure}[!t]
\vspace{0.1in}
\begin{subfigure}{.28\textwidth}
    \includegraphics[width=.99\linewidth, height=2cm]{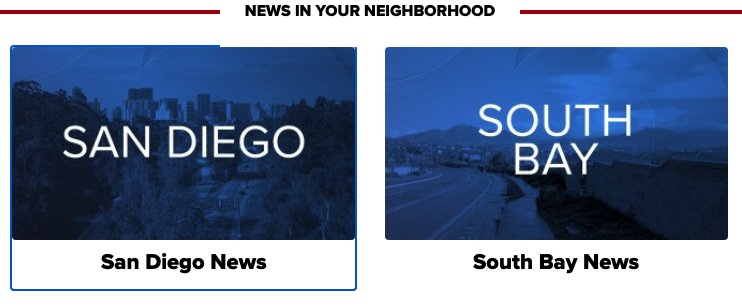}
    \caption{Focus Visible}
\end{subfigure}
\begin{subfigure}{.20\textwidth}
    \includegraphics[width=.90\linewidth,height=2cm]{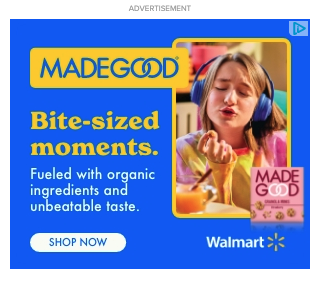}
    \caption{Focus Not Visible}
\end{subfigure}

\caption{The blue highlighted border shows the focus on the San Diego news block in (a), but it is not visible on the ad element in (b).} 
\label{fig:rq2-focus}
\vspace{-0.2in}
\end{figure} 
\subsection{RQ2: Common Accessibility Violations in Advertisements}
To further understand the reason behind the increase in violations caused by the ad elements on websites, we investigate the different types of violations, when these violations occur, and how common those are in ad elements.  
\lstdefinestyle{base}{
language=html,
moredelim=**[is][\color{red}]{@}{@},
moredelim=**[is][\color{darkgreen}]{~}{~},
numbers=left,
numberstyle=\tiny,
stepnumber=1,
numbersep=5pt,
escapechar=`
}
\begin{lstlisting}[caption={The corresponding {\tt <a>} tag for the aforementioned sponsored ad does not comply with WCAG on input criteria.},language=html, label={lst:on-input}, style=base]
<a id="aw0" target="_blank" href="https://googleads.g.doubleclick.net/pcs/click?....." onfocus="ss('aw0')" onmousedown="st('aw0')" onmouseover="ss('aw0')" onclick="ha('aw0')">
        <img src="https://tpc.googlesyndication.com/simgad/2522640214132067527" border="0" width="292" height="30" alt="" class="img_ad">  
</a>
\end{lstlisting}
\subsubsection{Prevalent Violations in Ad Elements (\roa)}
This analysis provides insights into design-specific violations within ad elements and their potential implications.
Table \ref{table:rq2-violations} highlights the most prevalent violations we found in ads across 5K websites.
Note that these ad elements with violations are rendered at runtime, preventing developers from addressing them during deployment.

\noindent\textbf{Focus Visible} is one of the most common violations. Focus visible results enforce that keyboard focus indicators are visible when styling elements like links or ad elements using CSS. 
In other words, website elements currently focused on should be highlighted. 
For example, on the website {\tt 10news.com}, we identified a Google Ads iframe that violates this rule, as depicted in Figure \ref{fig:rq2-focus}(b). 
In \ref{fig:rq2-focus}(a), the website developer has correctly implemented the focus visible feature by highlighting website cards with a blue border and thus adheres to focus visible. 
Under WCAG Success Criterion 2.4.7 (Focus Visible)\cite{WCAG-list}, focus indicators must always be visible, which benefits users with motor disabilities, visual impairments (including low vision), cognitive disabilities, and those relying on accessibility tools like screen readers.
We observe that 54.1\% of the website ad elements have not implemented focus visible.
%

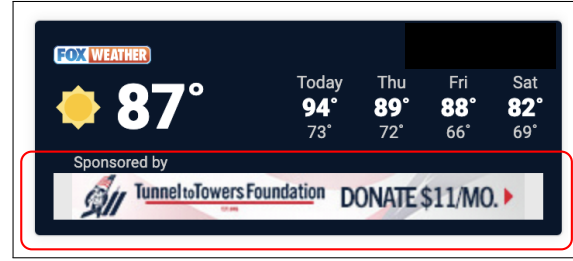
\begin{figure}[t]
    \centering
      \fbox{\includegraphics[width=0.40\textwidth]{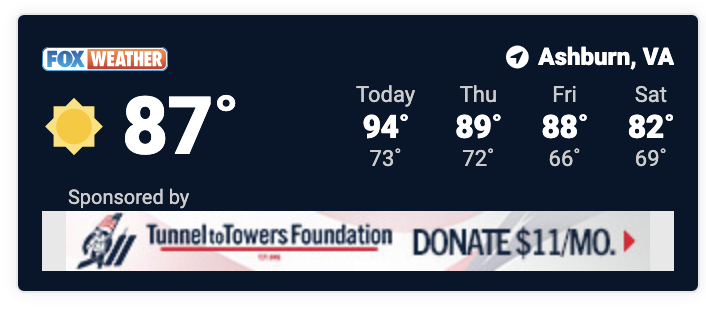}}
      \begin{tikzpicture}[overlay]
          \fill[black] (-2.4,2.4) rectangle (-0.4,3.);
          \draw[red,thick,rounded corners] (-7.5,0) rectangle (-0.18,1.3);
    \end{tikzpicture}
        \caption{The Fox News weather widget, featuring a sponsored ad element highlighted in a red box, does not comply with WCAG on input criteria.}
        \label{fig:rq2-on-input}
        \vspace{-0.3in}
\end{figure}

\begin{figure*}[!t]
\begin{subfigure}{.33\textwidth}
    \begin{tikzpicture}
\begin{axis}[
    ylabel={Proportion of Websites},
    legend pos=south east,
    ymin=0.7,
    ymax=1,
    xmin=0,
    xmax=6,
    grid=both,
    line width=1pt,
    axis line style={thick},
    width=1*\textwidth,
    height=4.5cm,
    legend style={font=\footnotesize} 
]

\addplot[color=red, dashed] coordinates {
(0.000, 0.836)
(1.000, 0.952)
(2.000, 0.971)
(3.000, 0.977)
(4.000, 0.993)
(5.000, 0.996)
(6.000, 0.996)
(7.000, 0.998)
(8.000, 0.999)
(10.000, 0.999)
(11.000, 0.999)
(12.000, 1.000)
(16.000, 1.000)
(18.000, 1.000)
};
\addplot[color=red] coordinates {
(0.000, 0.917)
(1.000, 0.990)
(2.000, 0.994)
(3.000, 0.997)
(4.000, 0.998)
(5.000, 0.999)
(6.000, 0.999)
(7.000, 0.999)
(8.000, 0.999)
(11.000, 1.000)
(14.000, 1.000)
};


\end{axis}
\end{tikzpicture}
    \vspace{-.4in}
    \caption{Missing Captions}
    \label{fig:normal}
\end{subfigure}
\hspace{-.2in} 
\begin{subfigure}{.33\textwidth}
    \begin{tikzpicture}
\begin{axis}[
    legend pos=south east,
    ymin=0.7,
    ymax=1,
    xmin=0,
    xmax=25,
    grid=both,
    line width=1pt,
    axis line style={thick},
    width=1*\textwidth,
    height=4.5cm,
]

\addplot[color=red, dashed] coordinates {
(0.000, 0.729)
(1.000, 0.802)
(2.000, 0.860)
(3.000, 0.889)
(4.000, 0.904)
(5.000, 0.915)
(6.000, 0.925)
(7.000, 0.927)
(8.000, 0.930)
(9.000, 0.931)
(10.000, 0.937)
(11.000, 0.972)
(12.000, 0.976)
(13.000, 0.980)
(14.000, 0.983)
(15.000, 0.991)
(16.000, 0.992)
(17.000, 0.993)
(18.000, 0.994)
(19.000, 0.995)
(20.000, 0.995)
(22.000, 0.995)
(23.000, 0.998)
(24.000, 0.998)
(29.000, 0.999)
(30.000, 0.999)
(34.000, 1.000)
(41.000, 1.000)
};
\addplot[color=red] coordinates {
(0.000, 0.809)
(1.000, 0.872)
(2.000, 0.933)
(3.000, 0.959)
(4.000, 0.974)
(5.000, 0.981)
(6.000, 0.987)
(7.000, 0.988)
(8.000, 0.990)
(9.000, 0.991)
(10.000, 0.993)
(11.000, 0.994)
(12.000, 0.995)
(13.000, 0.995)
(14.000, 0.996)
(15.000, 0.997)
(16.000, 0.997)
(18.000, 0.998)
(19.000, 0.998)
(22.000, 0.998)
(23.000, 0.999)
(24.000, 0.999)
(29.000, 0.999)
(30.000, 1.000)
(41.000, 1.000)
};


\end{axis}
\end{tikzpicture}
    \vspace{-.4in}
    \caption{Focus-order}
\end{subfigure}
\hspace{-.2in} 
\begin{subfigure}{.33\textwidth}
    \begin{tikzpicture}
\begin{axis}[
    legend pos=south east,
    ymin=0.4,
    ymax=1,
    xmin=0,
    xmax=12,
    grid=both,
    line width=1pt,
    axis line style={thick},
    width=1*\textwidth,
    height=4.5cm,
    legend style={font=\scriptsize}
]

\addplot[color=red, dashed] coordinates {
(0.000, 0.547)
(1.000, 0.705)
(2.000, 0.774)
(3.000, 0.838)
(4.000, 0.885)
(5.000, 0.914)
(6.000, 0.939)
(7.000, 0.952)
(8.000, 0.964)
(9.000, 0.972)
(10.000, 0.978)
(11.000, 0.983)
(12.000, 0.986)
(13.000, 0.990)
(14.000, 0.991)
(15.000, 0.993)
(16.000, 0.994)
(17.000, 0.995)
(18.000, 0.997)
(19.000, 0.997)
(20.000, 0.998)
(21.000, 0.999)
(22.000, 0.999)
(24.000, 0.999)
(27.000, 0.999)
(28.000, 1.000)
(30.000, 1.000)
};
\addplot[color=red] coordinates {
(0.000, 0.799)
(1.000, 0.969)
(2.000, 0.982)
(3.000, 0.998)
(4.000, 0.999)
(5.000, 0.999)
(9.000, 0.999)
(10.000, 1.000)
(12.000, 1.000)
(13.000, 1.000)
};

\legend{Ad elements, Website elements}

\end{axis}
\end{tikzpicture}
    \vspace{-.4in}
    \caption{Language of Page}
\end{subfigure}
\caption{Violations are predominantly found in ad elements (\roa) rather than in website elements (\rwoa).} 
\vspace{-4ex}
\label{fig:rq2-cdf}
\end{figure*}
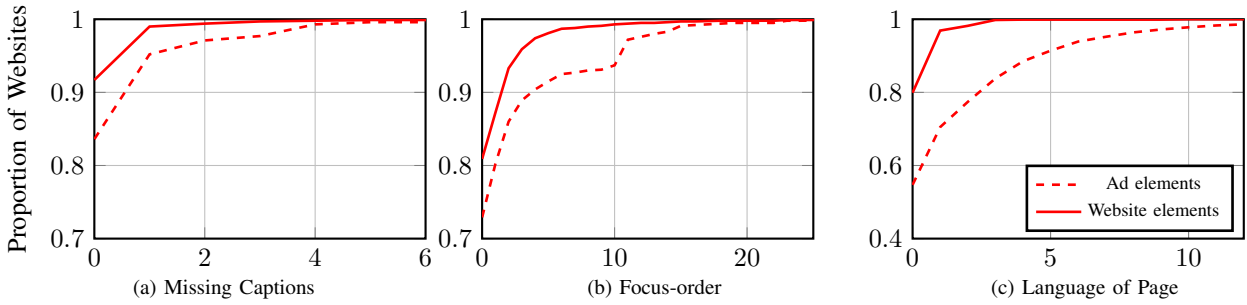
\noindent\textbf{In Input} is the second most common violation in ad elements that fails to inform users when their input action will change the context i.e., navigating to a different page or opening a new window. 
This requirement is mandated by WCAG Success Criterion 3.2.2 (On Input) \cite{WCAG-list}. 
According to this criterion, user actions on an interface component should not automatically change the context without prior notification. Violating "on input" causes confusion or disorientation from unexpected context changes.
%
We observe this violation on 47.1\% of the websites' ad elements.
For example, on the website {\tt foxnews.com}, we identified an ad that violates this rule, as shown in Figure \ref{fig:rq2-on-input}.
The red-highlighted ad has associated {\tt onclick}, {\tt onmouseover}, {\tt onmousedown}, and {\tt onfocus} events, as shown in Listing \ref{lst:on-input}.
These events open a new window (target="\_blank"), but there is no indication in the link text or an accessible tag like {\tt aria-label} to inform users that a new window will be opened. 
It is mandatory to inform users about changes in context to prevent disorientation.
\begin{tcolorbox}[colback=blue!5!white]
 {
    {\bf \em Finding.} Overall, we observed 28 different types of violations, with the most prevalent being Focus Visible (54\%), and In Input (47\%) across various websites. 
 }
\end{tcolorbox}

\subsubsection{Violations in Ad Elements (\roa) vs. Non-ads Website Elements \rwoa}
This analysis highlights the accessibility violations that are specific to ad elements (\roa) but not quite prevalent in non-ad elements (\rwoa). 
We compare violations from over 99K ad elements and 331K non-ad elements.
Figure \ref{fig:rq2-cdf}(a-c) shows the distribution of websites with the three most commonly found violations in ad elements (\roa), which are not common in non-ads website elements (\rwoa).

\noindent\textbf{Missing Captions} (WCAG 1.2.4 \cite{WCAG-list}) is one of the most common accessibility violations found more in ad elements  (\roa) than standard website elements  (\rwoa). 
%
This criterion requires that pre-recorded and live audio content in synchronized media (such as videos) includes captions.
Captions are essential for users who are deaf, as they provide a text alternative to spoken dialogue and important audio information.
On average, 16.4\% of ad elements (\roa) contain at least one instance of this violation, whereas only 8.3\% of website elements (\rwoa) exhibit the same violation. 
In other words, ad elements (\roa) are 2$\times$ more likely to be missing captions than non-ads elements (\rwoa) on websites.

\noindent\textbf{Focus Order} (WCAG 2.4.3 \cite{WCAG-list}) is found to be more prevalent in ad elements (\roa) than non-ads website elements (\rwoa).  
This criterion requires that the navigation order of interactive elements on a website is logical and intuitive, ensuring users can navigate through content in a predictable manner. 
Proper focus order is crucial for users who rely on keyboard navigation, such as individuals with motor impairments or those using assistive technologies like screen readers. 
Without a logical focus order, users may struggle to understand the structure of the page, miss important information, or find it difficult to interact with the content. 
On average, 27.1\% of ad elements (\roa) contain at least one instance of this violation, whereas only 19.1\% of website elements (\rwoa) exhibit the same violation.

\noindent\textbf{Language of Page}  (WCAG 3.1.1 \cite{WCAG-list}) is another accessibility violation that is found more in ad elements than non-ad website elements. 
This criterion requires that the primary language of a website element is clearly defined in the HTML. 
Declaring the language is essential for users who rely on assistive technologies, such as screen readers, which use this information to provide accurate pronunciation and interpretation of the content. 
Without the language of the page specified, users with visual impairments or cognitive disabilities may have difficulty understanding the content, leading to a less accessible and non-inclusive website experience. 
On average, 45.3\% of ad elements (\roa) contain at least one instance of this violation, whereas only 2.1\% of website elements (\rwoa) exhibit the same violation.  
Thus, ad elements (\roa) are 21$\times$ more likely to have this violation than non-ad elements (\rwoa).  
This significant difference points to the special ad accessibility case where website developers are not able to enforce language constraints on the ads served on their website due to the user-specific and dynamic nature of websites.  
\begin{tcolorbox}[colback=blue!5!white]
 {
    {\bf \em Finding.} Overall, we observed seven violations more common in ad elements (\roa) than in website elements (\rwoa): missing captions, incorrect focus order, undefined page language, insufficient text spacing, hover content violations, keyboard accessibility issues, and focus-related violations.
 }
\end{tcolorbox}

\begin{table*}[t]
\small
    \centering
    \scalebox{0.9}{\begin{tabularx}{\textwidth}{|p{1.8cm}|p{1.4cm}|l|l|l|X|l|}
    \hline
    \textbf{Domain} & \textbf{Company} & \textbf{Websites}& \textbf{Violations} & \textbf{WCAG } & \textbf{Description} & \textbf{Count} \\
    \hline
    \multirow{3}{*}{taboola.com} & \multirow{3}{*}{Taboola} & \multirow{3}{*}{411}& \multirow{3}{*}{17,522} & 1.1.1  & All non-text content have text alternatives to ensure accessibility for users who rely on assistive technologies. & 5,699 \\
    \cline{5-7}
    &&&& 3.2.2  & User inputs should not trigger unexpected changes or disruptions in content. &  5,057 \\
    \cline{5-7}
    &&&& 2.4.7& Any interactive element has a visible indicator of focus to help users navigate and interact with content. & 4,658\\
    \hline
    \multirow{3}{*}{doubleclick.net} & \multirow{3}{*}{Google} & \multirow{3}{*}{1,577} & \multirow{3}{*}{10,108}& 3.2.2  & User inputs should not trigger unexpected changes or disruptions in content. & 3,369 \\
    \cline{5-7}
    &&&& 2.4.4  & The purpose of a link is clear from its context, ensuring users understand where the link will take them. & 3,344 \\
    \cline{5-7}
    &&&& 2.4.7  & Any interactive element has a visible indicator of focus to help users navigate and interact with content. & 1,675\\
    \hline

    \multirow{3}{*}{revcontent.com} & \multirow{3}{*}{RevContent} & \multirow{3}{*}{325}& \multirow{3}{*}{3,204} & 3.2.2 & User inputs should not trigger unexpected changes or disruptions in content. & 1,375 \\
    \cline{5-7}
    &&&& 2.4.1  & Users can bypass repetitive content, such as navigation menus, to access the main content more efficiently. & 1,113\\
    \cline{5-7}
    &&&& 1.2.1  & Text alternatives/captions for prerecorded audio and video content. & 232\\
    \hline
    \multirow{3}{*}{\parbox{1.5cm}{googlesyndica\\tion.com}} & \multirow{3}{*}{Google} & \multirow{3}{*}{1,021}& \multirow{3}{*}{2,339} & 2.4.7  & Any interactive element has a visible indicator of focus to help users navigate and interact with content. & 1,844 \\
    \cline{5-7}
    &&&& 1.1.1 & All non-text content have text alternatives to ensure accessibility for users who rely on assistive technologies. &  410 \\
    \cline{5-7}
    &&&& 4.1.2  &  User interface elements have proper names, roles, and values to ensure they are accessible and understandable by assistive technologies. & 25\\
    \cline{5-7}
    \hline
    \multirow{3}{*}{google.com} & \multirow{3}{*}{Google} & \multirow{3}{*}{365}& \multirow{3}{*}{2,209} & 3.2.2 & User inputs should not trigger unexpected changes or disruptions in content. & 569 \\
    \cline{5-7}
    &&&& 4.1.2 &  User interface elements have proper names, roles, and values to ensure they are accessible and understandable by assistive technologies. &  516 \\
    \cline{5-7}
    &&&& 2.4.7  & Any interactive element has a visible indicator of focus to help users navigate and interact with content. & 504\\
    \cline{5-7}
    \hline
\end{tabularx}}
    \caption{Top five ad-serving technologies with the highest number of violations.}
    \label{table:rq3-adv-violations}
    \vspace{-0.25in}
\end{table*}
\subsection{RQ3: Most Prevalent Ad-Serving Technologies on Websites with Ad Serving Violations}
We investigate the ad-serving technologies that deliver ads on websites with the most accessibility violations. 
This analysis aims to guide website developers in selecting ad-serving technologies that comply with WCAG guidelines. 
Additionally, it encourages ad-serving technologies to prioritize accessibility considerations before delivering ads on websites.
Table \ref{table:rq3-adv-violations} lists the top ad-serving technologies prevalent on websites with the highest number of violations in ad elements (\roa). 

\noindent\textbf{Taboola}  is a content discovery platform that promotes personalized content recommendations across websites. 
In our investigation, it is found on 8.2\% of the inspected websites with a total of 18K violations.
Common ad violation is related to guideline 1.1.1 \cite{WCAG-list}, which mandates that all content must have a text alternative for non-text content, found in 5,699 ads.
Another common violation is related to guideline 3.2.2 \cite{WCAG-list}, which mandates providing adequate feedback when an action opens a new window or changes the context.
This guideline is violated in 5,057 ads. 
Lastly, guideline 2.4.7 \cite{WCAG-list}, which requires visible focus indicators for interface components to assist users navigating with keyboards or alternative devices, is violated in 4,658 ads served by Taboola.

\noindent\textbf{DoubleClick} 
%
is Google's ad management platform, optimizing digital ad campaigns. 
In our investigation, it is found on 31.5\% of inspected websites with a total of 10K violations.
Common ad violations include not informing users when their input action will open a new window or change context (Guideline 3.2.2) \cite{WCAG-list}, which is found in 3,369 ads.
Another common DoubleClick violation is related to guideline 2.4.4 \cite{WCAG-list}, which ensures links have clear, context-independent purposes,  benefiting screen reader users. 
This guideline is violated in 3,344 ads.
Lastly, Guideline 2.4.7 \cite{WCAG-list} mandates visible focus indicators for interface components, aiding users navigating with keyboards or alternative devices, and is violated in 1,675 ads served by DoubleClick.

\noindent\textbf{RevContent} is an ad-serving technology that delivers personalized ad recommendations across various websites. 
In our investigation, it is found on 6.5\% of inspected websites with a total of 3K violations.
Common ad violations include not informing users when their input action will open a new window or change context (Guideline 3.2.2) \cite{WCAG-list}, which is found in 1,375 ads.
Another common violation is related to guideline 2.4.1 \cite{WCAG-list}, which ensures that all functionality is accessible from a keyboard.
This guideline is violated in 1,113 ads.
Additionally, guideline 1.2.1 \cite{WCAG-list}, which requires that all multimedia content has a synchronized captioning or a transcript, is violated in 232 ads served by RevContent.



\begin{tcolorbox}[colback=blue!5!white]
 {
    {\bf \em Finding.} Overall, 52\% of websites using Google's ad-serving technologies violate WCAG guidelines across over 14K ad elements. Taboola serves the most ad elements with accessibility violations.
 }
\end{tcolorbox}

\subsection{Misrepresentation of Alt Text in Advertisements}
\label{sec:misrepesentation}
Among the ads that do not lead to accessibility violation, we investigate the accessible information associated with ads that truly represents the intext and content of the ad. 
Prior work privacy implications of ads~\cite{book2013longitudinal, book2015empirical, Alrizah19IMCerrorsMisunderstandings,Le21anticvndss,Bashir:2018:TCC:3278532.3278573} and malvertising~\cite{sood2011malvertising, xing2015understanding, arrate2020malvertising} show that clickbait~\cite{li2024attention,ahmad2024review} and false ads~\cite{handler1929false} can lead to scams, phishing, and fraud.  
Analyzing alternate texts of otherwise accessible ads is important because it highlights how effective alternate texts are in effectively conveying the content of ads to users with disabilities.

We start by extracting {\tt <img>} tags with alt text from the ads' HTML and collecting their {\tt src} values, which contain the image URLs. Next, we use the CLIP (Contrastive Language–Image Pretraining) model~\cite{OpenAI-Clip} to assess the relevance between images and their alt text. CLIP encodes images and text into a shared embedding space, allowing us to compute the cosine similarity between them. We classify alt text as redundant if the cosine similarity is below 0.5, which is the optimal threshold reported in previous research \cite{ahmad2024scs}. This approach efficiently evaluates the semantic alignment between visual and textual data.

\begin{figure}[!t]
\vspace{0.1in}
\begin{subfigure}{.23\textwidth}
    \includegraphics[width=.99\linewidth, height=2cm]{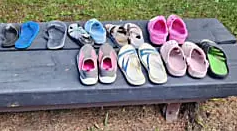}
    \caption{Alt-text: Foxes accused of stealing Crocs, sandals from campers after many pairs found at den.}
\end{subfigure}
\begin{subfigure}{.23\textwidth}
    \includegraphics[width=.99\linewidth, height=2cm]{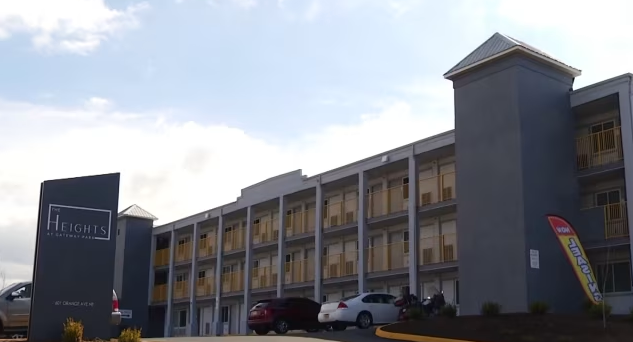}
    \caption{Alt-text: Distressed hotels turned into apartments for affordable housing.}
\end{subfigure}

\caption{Misrepresentation of alt-text in image tags of ads.} 
\vspace{-0.3in}
\label{fig:rq5-misrepresentation}
\end{figure}

Overall, across 5K websites,  we find 9,827 ad images with alt text.
Of these, 27.1\% of ads have a cosine similarity score below 0.5, indicating a low level of similarity between the text and the ad image.
For example, Figure \ref{fig:rq5-misrepresentation}(a) shows an image of shoes on a bench, served by Taboola on {\tt wsfa.com}, but the alt text describes foxes stealing footwear. This alt text mimics a news headline to confuse users, with a cosine similarity score of 0.33.
Similarly, Figure \ref{fig:rq5-misrepresentation}(b) shows a hotel image on {\tt clickorlando.com}, but the alt text describes hotels turned into apartments for affordable housing. This alt text also resembles a news headline,  with a cosine similarity score of 0.28.
%
The alt texts of these ads do not disclose that they are ads and not the content served by the website. 
%
Users may only realize it is an ad after performing actions like pressing enter to open the link, only to be redirected to a page outside the original domain. 
Malicious and false ads represented as news in alt text can redirect to scam and phishing websites, compromising the safety of users with disabilities who struggle to assess a website's trustworthiness. 
\begin{tcolorbox}[colback=blue!5!white]
 {
    {\bf \em Finding.} Overall, 27\% of the nearly 10K ads have alt text that misrepresents ad image's content and intent.
 }
\end{tcolorbox}

\noindent\textbf{Privacy implication of misleading advertisements.} We manually inspect the consequences of interacting with these misleading ads to determine if any user identifiers are collected for invasive tracking {\tt onhover}, {\tt onmouseover}, or when the user presses enter (for tabbable access) on these ads. 
These three events are key resources for assistive technologies like screen readers to collect accessible information from website elements and deliver it to users using them

Prior research has shown that network traffic resulting from interacting with ads may contain user identifiers in headers \cite{rieder2024beyond, bujlow2015web}, cookies \cite{munir2023cookiegraph, amjad2024blocking}, and parameters \cite{munir2023purl}. 
These identifiers can be used to trace a user's browsing activity on the web, violating their privacy. 
Lists maintained by ad-blockers identify specific headers\cite{Headers}, cookies\cite{CookieList}, and parameters\cite{AdguardBlockParam} as tracking identifiers for users. 
We compare our network traffic from manually interacting with ads on the top 50 news websites against these lists to identify any tracking of user identifiers. 
Our results show that on 94\% of websites, at least one tracking identifier is sent to an external server by hovering, and on 98\% of websites, at least one tracking identifier is sent to an external server by pressing enter on the ads. 
Both events are essential actions that screen readers perform to read and announce website elements' content. None of the actions are necessary to access ads for regular users. 
%
\begin{tcolorbox}[colback=blue!5!white]
 {
    {\bf \em Finding.} On 94\% of websites, hovering sends tracking data to external servers, forcing users with disabilities to compromise privacy to navigate ad content, unlike regular users.
 }
\end{tcolorbox}
\section{Discussion and Takeaways}
\label{sec:discussion}
Based on our empirical investigation, we provides insights for website developers and ad-serving technologies to address ad accessibility violations and highlights future research avenues for improving the website experience for users with disabilities.

\noindent\textbf{Accessibility-promoting ad integration for website developers.}
Ad integration is crucial for ad-driven websites. Developers must ensure high accessibility and WCAG compliance, which can be a regulatory requirement\cite{WCAG-21}. Once published, the developer cannot modify the website to address ad-related accessibility issues since new ads may introduce different violations. Developers can ensure two key aspects to improve ad accessibility during integration.
%
First, they can inject JavaScript to enforce CSS rules (using {\tt !important}) to implement basic checks on WCAG Type A violations. 
For instance, a script can be added to ensure all text elements have sufficient color contrast. Such a script can be triggered on {\tt onload} events to detect certain accessibility violations and address them appropriately, if possible.  
Some Type AA violations can also be addressed by injecting similar checks in the JavaScript, such as ensuring all interactive elements are keyboard accessible.
Secondly, developers can run accessibility checks on their websites post-deployment (similar to our analysis) to analyze ad elements and derive specific self-correcting strategies to address those concerns.

\noindent\textbf{Accessibility strategies for ad-serving technologies.}
Website developers integrate ads on their websites by leasing ad space to ad-serving technologies. 
These ad-serving technologies often enroll in programs such as the Acceptable Ads program \cite{Acceptable-ads} and adhere to guidelines set by the Interactive Advertising Bureau (IAB) \cite{IAB}.
The Acceptable Ads Program promotes non-intrusive ad formats that respect user experience, requiring adherence to specific display guidelines. 
The IAB sets comprehensive standards for digital advertising, including ad formats, measurement, and data privacy, ensuring compliance with industry best practices.
However, none of these current programs and IAB standards address the accessibility of ads, such as readability by screen readers or keyboard navigation. 
Including accessibility criteria in these guidelines would ensure that ads meet website accessibility standards. 
Updating IAB standards to cover accessibility requirements would support assistive technologies and enhance compliance with website accessibility guidelines.
When leasing an ad slot, the website developer is often provided with the option to enforce restrictions on the type of ads they want to display on their websites. Based on our findings, we advocate that this ad restriction should also include an option to host only WCAG-complaint ads. 

\noindent\textbf{User-side accessibility tools.}
Recent advancements in large language models (LLMs) can significantly benefit accessibility tools by addressing common accessibility violations automatically. 
For instance, when dealing with missing {\tt alt text} for images, LLMs can analyze the image's content and generate descriptive and contextually relevant {\tt alt text}. 
These models can provide accurate descriptions by leveraging advanced image recognition and natural language processing capabilities, making the content more accessible to users with visual impairments. 
However, hosting large-sized models in a browser extension is not feasible. Small, domain-specific language models are an active area of research, which can advance a browser's ability to auto-transform inaccessible ads into WCAG-complaint with minimal runtime cost.
\section{Threat to validity}
\label{sec:validity}
\noindent\textbf{Internal validity.}
Our analysis in Section~\ref{sec:results} involves collecting ad elements (\roa) by differencing website elements without ads (\rwoa) from those with ads (\rwa). 
However, various confounding factors may influence \roa.
For instance, some websites dynamically fetch different elements, including ads, on each visit.
Consequently, website elements loaded in \rwa might not appear in \rwoa. 
Other factors include changes in behavior due to the environment (\eg browser and host OS), visit time, and location.
To minimize this threat, we perform the two visits in quick succession and maintain consistent environments across different states, ensuring the same location, browser, and stateless crawls.

\noindent\textbf{External validity.} 
The IBM Equal Access Accessibility Checker \cite{IBM-accessibility} on different browsers might produce different accessibility violations due to the browser's built-in features, such as Safari's default blocking of certain ad-serving technologies \cite{Safari-ITP}. 
We conduct our experiments using the Chrome browser with a Chrome-based extension, which is the most widely used browser, to minimize this threat.

\noindent\textbf{Construct validity.}
We collect the website's elements at page load time and do not capture changes in accessibility violations triggered by user interactions such as scrolling and clicking. This is a general limitation of dynamic analysis that is challenging to overcome due to the numerous navigation and interaction scenarios on websites. 
\section{Related Work}
\label{sec:related-work}

Accessibility of websites and mobile applications has been extensively studied \cite{hackett2003accessibility, campoverde2023accessibility, mohammadi2024accessibility, agrawal2019evaluating, teixeira2021diversity, fok2022large, milne2014accessibility, ross2018examining}. 
Hackett et al. \cite{campoverde2023accessibility} used the Internet Archive’s Wayback Machine to demonstrate that websites have become progressively less accessible as their complexity has increased. 
Teixeira et al. \cite{teixeira2021diversity} conducted a systematic literature review focusing on website accessibility within the tourism industry. 
Similarly, Moreno et al. \cite{moreno2008guiding} proposed a method to improve website design for accessibility by addressing identified violations. 
Fok et al. \cite{fok2022large} conducted a large-scale analysis of mobile applications to investigate how inaccessible user interfaces are. 
Similarly, Milne et al. \cite{milne2014accessibility} and Ross et al. \cite{ross2018examining} found that missing or vague labels in mobile applications hinder screen readers from conveying information to visually impaired users.

 Digital ads on websites and mobile applications are mainly studied for privacy-invasive targeted advertising mechanisms (\eg \cite{bashir2016tracing, englehardt2016online, eslami2018communicating}), malicious ads spreading malware or click fraud and phishing attacks (\eg \cite{li2012knowing, rastogi2016these}), and deceptive or manipulative content (\eg \cite{zeng2020bad}).
However, there is very limited research specifically focused on the accessibility of online ads.
He et al. \cite{he2024tend} conducted an empirical study on the accessibility of mobile ads for blind users in Android apps, identifying significant challenges such as missing {\tt alt text} for images, complex gestures problematic for screen readers, and ads that disrupt app functionality. 
However, their analysis was limited to only 500 ad elements on mobile applications, so their insights are not applicable to website ads broadly.
Similarly, Thompson et al. \cite{thompson2001accessibility} manually examined {\tt alt text} for ad elements on 67 news websites, while Chiou et al. \cite{10.1145/3597503.3639229} automatically detected reflow accessibility violations in website advertisements for keyboard users. Both studies are small-scale and on limited accessibility violations. 
To date, the large-scale accessibility of website ads remains a significant and unresolved question.

Several tools\cite{kumar2021comparing,gleason2020twitter,zhang2021screen, prakash2024all} have been developed to address accessibility issues. 
Twitter A11y \cite{gleason2020twitter} is a browser extension designed to add alternative text to images on Twitter.
Zhang et al. \cite{zhang2021screen} created a machine-learning model that infers accessibility metadata from mobile app screens, improving iOS VoiceOver functionality and enhancing accessibility for existing apps. 
InstaFetch \cite{prakash2024all} is a browser extension that consolidates website content into a screen reader-friendly interface, allowing blind users to access information across multiple pages efficiently.
while these tools may improve the accessibility of general website content, it is yet to be measured if these fully address the accessibility violation caused by the ads.

\vspace{-0.1in}
\section{Conclusion}
\label{sec:conclusion}

Website ads are an integral part of a website and are essential for supporting free Internet; however, they often fail to meet basic accessibility standards.
This paper presents the first extensive study on the accessibility issues of website ads. Our research reveals that 67\% of websites show a noticeable increase in accessibility violations due to ads. We also discovered that popular ad-serving technologies do not adhere to the WCAG guidelines when displaying ads, and nearly one-quarter of ads that do comply with WCAG guidelines have misleading alt text.
Our findings provide insight into the challenges faced by website developers in integrating ads while maintaining accessibility. This research suggests ways for developers to enforce accessibility standards to lessen the impact of ads on overall website's accessibility.





\bibliographystyle{IEEEtran}
\bibliography{conference_101719}

\begin{thebibliography}{10}
\providecommand{\url}[1]{#1}
\csname url@samestyle\endcsname
\providecommand{\newblock}{\relax}
\providecommand{\bibinfo}[2]{#2}
\providecommand{\BIBentrySTDinterwordspacing}{\spaceskip=0pt\relax}
\providecommand{\BIBentryALTinterwordstretchfactor}{4}
\providecommand{\BIBentryALTinterwordspacing}{\spaceskip=\fontdimen2\font plus
\BIBentryALTinterwordstretchfactor\fontdimen3\font minus \fontdimen4\font\relax}
\providecommand{\BIBforeignlanguage}[2]{{%
\expandafter\ifx\csname l@#1\endcsname\relax
\typeout{** WARNING: IEEEtran.bst: No hyphenation pattern has been}%
\typeout{** loaded for the language `#1'. Using the pattern for}%
\typeout{** the default language instead.}%
\else
\language=\csname l@#1\endcsname
\fi
#2}}
\providecommand{\BIBdecl}{\relax}
\BIBdecl

\bibitem{brown1992assistive}
C.~Brown, ``Assistive technology computers and persons with disabilities,'' \emph{Communications of the ACM}, vol.~35, no.~5, pp. 36--45, 1992.

\bibitem{WCAG-21}
``Wcag-21,'' \url{https://www.w3.org/TR/WCAG21/}, 2024.

\bibitem{kumar2021comparing}
S.~Kumar, J.~Shree~DV, and P.~Biswas, ``Comparing ten wcag tools for accessibility evaluation of websites,'' \emph{Technology and Disability}, vol.~33, no.~3, pp. 163--185, 2021.

\bibitem{gleason2020twitter}
C.~Gleason, A.~Pavel, E.~McCamey, C.~Low, P.~Carrington, K.~M. Kitani, and J.~P. Bigham, ``Twitter a11y: A browser extension to make twitter images accessible,'' in \emph{Proceedings of the 2020 chi conference on human factors in computing systems}, 2020, pp. 1--12.

\bibitem{zhang2021screen}
X.~Zhang, L.~De~Greef, A.~Swearngin, S.~White, K.~Murray, L.~Yu, Q.~Shan, J.~Nichols, J.~Wu, C.~Fleizach \emph{et~al.}, ``Screen recognition: Creating accessibility metadata for mobile applications from pixels,'' in \emph{Proceedings of the 2021 CHI Conference on Human Factors in Computing Systems}, 2021, pp. 1--15.

\bibitem{prakash2024all}
Y.~Prakash, A.~K. Nayak, M.~Sunkara, S.~Jayarathna, H.-N. Lee, and V.~Ashok, ``All in one place: Ensuring usable access to online shopping items for blind users,'' \emph{Proceedings of the ACM on Human-Computer Interaction}, vol.~8, no. EICS, pp. 1--25, 2024.

\bibitem{evans2009online}
D.~S. Evans, ``The online advertising industry: Economics, evolution, and privacy,'' \emph{Journal of economic perspectives}, vol.~23, no.~3, pp. 37--60, 2009.

\bibitem{Taboola}
``Taboola,'' \url{https://www.taboola.com/}, 2024.

\bibitem{Google-lighthouse}
``Google-lighthouse,'' \url{https://developer.chrome.com/docs/lighthouse/overview}, 2024.

\bibitem{bashir2016tracing}
M.~A. Bashir, S.~Arshad, W.~Robertson, and C.~Wilson, ``Tracing information flows between ad exchanges using retargeted ads,'' in \emph{25th USENIX Security Symposium (USENIX Security 16)}, 2016, pp. 481--496.

\bibitem{englehardt2016online}
S.~Englehardt and A.~Narayanan, ``Online tracking: A 1-million-site measurement and analysis,'' in \emph{Proceedings of the 2016 ACM SIGSAC conference on computer and communications security}, 2016, pp. 1388--1401.

\bibitem{eslami2018communicating}
M.~Eslami, S.~R. Krishna~Kumaran, C.~Sandvig, and K.~Karahalios, ``Communicating algorithmic process in online behavioral advertising,'' in \emph{Proceedings of the 2018 CHI conference on human factors in computing systems}, 2018, pp. 1--13.

\bibitem{li2012knowing}
Z.~Li, K.~Zhang, Y.~Xie, F.~Yu, and X.~Wang, ``Knowing your enemy: understanding and detecting malicious web advertising,'' in \emph{Proceedings of the 2012 ACM conference on Computer and communications security}, 2012, pp. 674--686.

\bibitem{rastogi2016these}
V.~Rastogi, R.~Shao, Y.~Chen, X.~Pan, S.~Zou, and R.~D. Riley, ``Are these ads safe: Detecting hidden attacks through the mobile app-web interfaces.'' in \emph{NDSS}, 2016.

\bibitem{hackett2003accessibility}
S.~Hackett, B.~Parmanto, and X.~Zeng, ``Accessibility of internet websites through time,'' in \emph{Proceedings of the 6th International ACM SIGACCESS Conference on Computers and Accessibility}, 2003, pp. 32--39.

\bibitem{campoverde2023accessibility}
M.~Campoverde-Molina, S.~Luj{\'a}n-Mora, and L.~Valverde, ``Accessibility of university websites worldwide: a systematic literature review,'' \emph{Universal Access in the Information Society}, vol.~22, no.~1, pp. 133--168, 2023.

\bibitem{mohammadi2024accessibility}
M.~K. Mohammadi, V.~Esichaikul, and A.~Mohammadi, ``Accessibility and usability evaluation of university websites in afghanistan: a comparison between public and private universities,'' \emph{Universal Access in the Information Society}, vol.~23, no.~2, pp. 955--974, 2024.

\bibitem{agrawal2019evaluating}
G.~Agrawal, D.~Kumar, M.~Singh, and D.~Dani, ``Evaluating accessibility and usability of airline websites,'' in \emph{Advances in Computing and Data Sciences: Third International Conference, ICACDS 2019, Ghaziabad, India, April 12--13, 2019, Revised Selected Papers, Part I 3}.\hskip 1em plus 0.5em minus 0.4em\relax Springer, 2019, pp. 392--402.

\bibitem{teixeira2021diversity}
P.~Teixeira, C.~Eus{\'e}bio, and L.~Teixeira, ``Diversity of web accessibility in tourism: Evidence based on a literature review,'' \emph{Technology and Disability}, vol.~33, no.~4, pp. 253--272, 2021.

\bibitem{fok2022large}
R.~Fok, M.~Zhong, A.~S. Ross, J.~Fogarty, and J.~O. Wobbrock, ``A large-scale longitudinal analysis of missing label accessibility failures in android apps,'' in \emph{Proceedings of the 2022 CHI Conference on Human Factors in Computing Systems}, 2022, pp. 1--16.

\bibitem{milne2014accessibility}
L.~R. Milne, C.~L. Bennett, and R.~E. Ladner, ``The accessibility of mobile health sensors for blind users.(dec. 2014),'' 2014.

\bibitem{ross2018examining}
A.~S. Ross, X.~Zhang, J.~Fogarty, and J.~O. Wobbrock, ``Examining image-based button labeling for accessibility in android apps through large-scale analysis,'' in \emph{Proceedings of the 20th International ACM SIGACCESS Conference on Computers and Accessibility}, 2018, pp. 119--130.

\bibitem{he2024tend}
Z.~He, S.~F. Huq, and S.~Malek, ``" i tend to view ads almost like a pestilence": On the accessibility implications of mobile ads for blind users,'' in \emph{Proceedings of the IEEE/ACM 46th International Conference on Software Engineering}, 2024, pp. 1--13.

\bibitem{thompson2001accessibility}
D.~Thompson and B.~Wassmuth, ``Accessibility of online advertising: a content analysis of alternative text for banner ad images in online newspapers,'' \emph{Disability Studies Quarterly}, vol.~21, no.~2, 2001.

\bibitem{10.1145/3597503.3639229}
\BIBentryALTinterwordspacing
P.~T. Chiou, R.~Winn, A.~S. Alotaibi, and W.~G.~J. Halfond, ``Automatically detecting reflow accessibility issues in responsive web pages,'' in \emph{Proceedings of the IEEE/ACM 46th International Conference on Software Engineering}, ser. ICSE '24.\hskip 1em plus 0.5em minus 0.4em\relax New York, NY, USA: Association for Computing Machinery, 2024. [Online]. Available: \url{https://doi.org/10.1145/3597503.3639229}
\BIBentrySTDinterwordspacing

\bibitem{OpenAI-Clip}
``Openai-clip,'' \url{https://openai.com/index/clip/}, 2024.

\bibitem{alaskanews}
``alaskanews,'' \url{https://www.alaskasnewssource.com/}, 2024.

\bibitem{Google-ads}
``Google-ads,'' \url{https://ads.google.com/home/}, 2024.

\bibitem{IBM-accessibility}
``Ibm-accessibility,'' \url{https://chromewebstore.google.com/detail/ibm-equal-access-accessib/lkcagbfjnkomcinoddgooolagloogehp?hl=en-US}, 2024.

\bibitem{uBlock-origin}
``ublock-origin,'' \url{https://chromewebstore.google.com/detail/ublock-origin/cjpalhdlnbpafiamejdnhcphjbkeiagm?hl=en}, 2024.

\bibitem{WCAG-list}
``Wcag-guidelines,'' \url{https://www.w3.org/TR/2023/REC-WCAG21-20230921/}, 2024.

\bibitem{Chrome-vox}
``Chrome-vox,'' \url{https://chromewebstore.google.com/detail/screen-reader/kgejglhpjiefppelpmljglcjbhoiplfn}, 2024.

\bibitem{Similar-web}
``Similar-web,'' \url{https://www.similarweb.com/top-websites/news-and-media/}, 2024.

\bibitem{zeng2020bad}
E.~Zeng, T.~Kohno, and F.~Roesner, ``Bad news: Clickbait and deceptive ads on news and misinformation websites,'' in \emph{Workshop on Technology and Consumer Protection}, 2020, pp. 1--11.

\bibitem{easylist}
``easylist,'' \url{https://github.com/easylist/easylist}, 2024.

\bibitem{easyprivacy}
``easyprivacy,'' \url{https://github.com/easylist/easylist/tree/master/easyprivacy}, 2024.

\bibitem{Adblockplus}
``adblockplus,'' \url{https://adblockplus.org/}, 2024.

\bibitem{adguard}
``adguard,'' \url{https://adguard.com/en/welcome.html}, 2024.

\bibitem{castell2022demystifying}
I.~Castell-Uroz, R.~Sanz-Garc{\'\i}a, J.~Sol{\'e}-Pareta, and P.~Barlet-Ros, ``Demystifying content-blockers: Measuring their impact on performance and quality of experience,'' \emph{IEEE Transactions on Network and Service Management}, vol.~19, no.~3, pp. 3562--3573, 2022.

\bibitem{Axe}
``axe,'' \url{https://www.deque.com/axe/}, 2024.

\bibitem{Wave}
``wave,'' \url{https://wave.webaim.org/}, 2024.

\bibitem{Seo-accessibility}
``Seo-accessibility,'' \url{https://adasitecompliance.com/website-accessibility-seo-impact/}, 2024.

\bibitem{book2013longitudinal}
T.~Book, A.~Pridgen, and D.~S. Wallach, ``Longitudinal analysis of android ad library permissions,'' \emph{arXiv preprint arXiv:1303.0857}, 2013.

\bibitem{book2015empirical}
T.~Book and D.~S. Wallach, ``An empirical study of mobile ad targeting,'' \emph{arXiv preprint arXiv:1502.06577}, 2015.

\bibitem{Alrizah19IMCerrorsMisunderstandings}
M.~Alrizah, S.~Zhu, X.~Xing, and G.~Wang, ``{Errors, Misunderstandings, and Attacks: Analyzing the Crowdsourcing Process of Ad-blocking Systems},'' in \emph{ACM Internet Measurement Conference (IMC)}, 2019.

\bibitem{Le21anticvndss}
H.~Le, A.~Markopoulou, and Z.~Shafiq, ``Cv-inspector: Towards automating detection of adblock circumvention,'' in \emph{Network and Distributed System Security Symposium (NDSS)}, 2021.

\bibitem{Bashir:2018:TCC:3278532.3278573}
M.~A. Bashir, S.~Arshad, E.~Kirda, W.~Robertson, and C.~Wilson, ``How tracking companies circumvented ad blockers using websockets,'' in \emph{Proceedings of the Internet Measurement Conference (IMC)}, 2018.

\bibitem{sood2011malvertising}
A.~K. Sood and R.~J. Enbody, ``Malvertising--exploiting web advertising,'' \emph{Computer Fraud \& Security}, vol. 2011, no.~4, pp. 11--16, 2011.

\bibitem{xing2015understanding}
X.~Xing, W.~Meng, B.~Lee, U.~Weinsberg, A.~Sheth, R.~Perdisci, and W.~Lee, ``Understanding malvertising through ad-injecting browser extensions,'' in \emph{Proceedings of the 24th international conference on world wide web}, 2015, pp. 1286--1295.

\bibitem{arrate2020malvertising}
A.~Arrate, J.~Gonz{\'a}lez-Caba{\~n}as, {\'A}.~Cuevas, and R.~Cuevas, ``Malvertising in facebook: Analysis, quantification and solution,'' \emph{Electronics}, vol.~9, no.~8, p. 1332, 2020.

\bibitem{li2024attention}
X.~Li, J.~Zhou, H.~Xiang, and J.~Cao, ``Attention grabbing through forward reference: An erp study on clickbait and top news stories,'' \emph{International Journal of Human--Computer Interaction}, vol.~40, no.~11, pp. 3014--3029, 2024.

\bibitem{ahmad2024review}
A.~A. Ahmad~Azam, N.~F.~F. Hasbullah, N.~M. Abdul~Sookor, W.~Mohamad~Hanafi, M.~S. Hassan, and N.~A.~N. Ibrahim, ``A review of the effects of clickbait on online platforms among society,'' \emph{e-Journal of Media and Society (e-JOMS)}, vol.~7, no.~1, pp. 97--117, 2024.

\bibitem{handler1929false}
M.~Handler, ``False and misleading advertising,'' \emph{Yale LJ}, vol.~39, p.~22, 1929.

\bibitem{ahmad2024scs}
M.~Ahmad and M.~Mazzara, ``Scs-net: Sharpend cosine similarity based neural network for hyperspectral image classification,'' \emph{IEEE Geoscience and Remote Sensing Letters}, 2024.

\bibitem{rieder2024beyond}
W.~Rieder, P.~Raschke, and T.~Cory, ``Beyond the request: Harnessing http response headers for cross-browser web tracker classification in an imbalanced setting,'' \emph{arXiv preprint arXiv:2402.01240}, 2024.

\bibitem{bujlow2015web}
T.~Bujlow, V.~Carela-Espa{\~n}ol, J.~Sol{\'e}-Pareta, and P.~Barlet-Ros, ``Web tracking: Mechanisms, implications, and defenses,'' \emph{arXiv preprint arXiv:1507.07872}, 2015.

\bibitem{munir2023cookiegraph}
S.~Munir, S.~Siby, U.~Iqbal, S.~Englehardt, Z.~Shafiq, and C.~Troncoso, ``Cookiegraph: Understanding and detecting first-party tracking cookies,'' in \emph{Proceedings of the 2023 ACM SIGSAC Conference on Computer and Communications Security}, 2023, pp. 3490--3504.

\bibitem{amjad2024blocking}
A.~H. Amjad, S.~Munir, Z.~Shafiq, and M.~A. Gulzar, ``Blocking tracking javascript at the function granularity,'' \emph{arXiv preprint arXiv:2405.18385}, 2024.

\bibitem{munir2023purl}
S.~Munir, P.~Lee, U.~Iqbal, Z.~Shafiq, and S.~Siby, ``Purl: Safe and effective sanitization of link decoration,'' \emph{arXiv preprint arXiv:2308.03417}, 2023.

\bibitem{Headers}
``headers,'' \url{https://github.com/uBlockOrigin/uAssets/blob/master/filters/badlists.txt}, 2024.

\bibitem{CookieList}
``Cookielist,'' \url{https://raw.githubusercontent.com/cookiegraph/CookieGraph/main/Data/ublock_filterlist.txt}, 2024.

\bibitem{AdguardBlockParam}
``Adguardblockparam,'' \url{https://raw.githubusercontent.com/AdguardTeam/FiltersRegistry/master/filters/filter_17_TrackParam/filter.txt}, 2024.

\bibitem{Acceptable-ads}
``Acceptable-ads,'' \url{https://acceptableads.com/}, 2024.

\bibitem{IAB}
``Iab-ads,'' \url{https://www.iab.com/guidelines/iab-new-ad-portfolio}, 2024.

\bibitem{Safari-ITP}
``Safari-itp,'' \url{https://clearcode.cc/blog/intelligent-tracking-prevention/}, 2024.

\bibitem{moreno2008guiding}
L.~Moreno, P.~Mart{\'\i}nez, and B.~Ruiz, ``Guiding accessibility issues in the design of websites,'' in \emph{Proceedings of the 26th annual ACM international conference on Design of communication}, 2008, pp. 65--72.

\end{thebibliography}

\end{document}